\documentclass[english]{revtex4-2}
\usepackage[T1]{fontenc}
\usepackage[utf8]{inputenc}
\usepackage{refstyle}
\usepackage{mathtools}
\usepackage{amsmath}
\usepackage{graphicx}

\makeatletter

\AtBeginDocument{\providecommand\secref[1]{\ref{sec:#1}}}
\AtBeginDocument{\providecommand\subsecref[1]{\ref{subsec:#1}}}
\AtBeginDocument{\providecommand\figref[1]{\ref{fig:#1}}}
\AtBeginDocument{\providecommand\eqref[1]{\ref{eq:#1}}}
\AtBeginDocument{\providecommand\appref[1]{\ref{app:#1}}}
\newcommand{\lyxdot}{.}

\RS@ifundefined{subsecref}
  {\newref{subsec}{name = \RSsectxt}}
  {}
\RS@ifundefined{thmref}
  {\def\RSthmtxt{theorem~}\newref{thm}{name = \RSthmtxt}}
  {}
\RS@ifundefined{lemref}
  {\def\RSlemtxt{lemma~}\newref{lem}{name = \RSlemtxt}}
  {}

\usepackage{babel}
\usepackage{amssymb}
\usepackage{feynmp-auto}
\usepackage{MnSymbol}
\let\citet\cite

\usepackage{chngcntr}

\makeatletter
\newcommand{\appendixfigures}{
  \counterwithin{figure}{section}
  \renewcommand{\thefigure}{\thesection\arabic{figure}}
  \@addtoreset{figure}{section}
}
\makeatother

\makeatother

\usepackage{babel}
\begin{document}
\global\long\def\MIR{\mathrm{MIR}}%
\global\long\def\MI{{\cal I}}%
\global\long\def\dt{\mathrm{dt}}%
\global\long\def\skew{\gamma}%
\global\long\def\ms{\mathrm{ms}}%

\title{Diagrammatic expansion for the mutual-information rate in the realm
of limited statistics}
\author{Tobias Kühn$^{1}$}
\thanks{Contact: tobias.kuhn@inserm.fr, current address: Institut für Physiologie,
Universität Bern, Muesmattstr 27a, 3012 Bern, Switzerland}
\author{, Gabriel Mahuas$^{1,2}$, Ulisse Ferrari$^{1}$}
\affiliation{$^{1}$Institut de la Vision, Sorbonne Université, CNRS, INSERM, 17
rue Moreau, 75012, Paris, France}
\affiliation{$^{2}$Laboratoire de physique de l'École normale supérieure, CNRS,
PSL University, Sorbonne University, Université Paris-Cité, 24 rue
Lhomond, 75005 Paris, France}
\begin{abstract}
Neurons in sensory systems encode stimulus information into their
stochastic spiking response. The mutual information has been extensively
applied to these systems to quantify the neurons' capacity of transmitting
such information. Yet, while for discrete stimuli, like flashed images
or single tones, its computation is straightforward, for dynamical
stimuli it is necessary to compute a (mutual) information rate (MIR),
therefore integrating over the multiple temporal correlations which
characterize sensory systems. Previous methods are based on extensive
sampling of the neuronal response, require large amounts of data and
are therefore prone to biases and inaccuracy. Here, we develop Moba-MIRA
(moment-based mutual-information-rate approximation), a computational
method to estimate the mutual information rate. To derive Moba-MIRA,
we use Feynman diagrams to expand the mutual information to arbitrary
order in the correlations around the corresponding value for the empirical
spike count distributions of single bins. As a result, only the empirical
estimation of the pairwise correlations between time bins and the
single-bin entropies are required, without the need for the whole
joint probability distributions. We tested Moba-MIRA on synthetic
data generated with generalized linear models, and showed that it
requires only a few tens of stimulus repetitions to provide an accurate
estimate of the information rate. Finally, we applied it to ex-vivo
electrophysiological recordings of rats retina, obtaining rates ranging
between 5 to 20 bits per second, consistent with earlier estimates.
\end{abstract}
\maketitle
\let\clearpage\relax

\section{Introduction}

Neurons in sensory systems respond to stimulus presentation by changes
in their electrical potential, often with the emission of spikes,
which are then sent to downstream areas for further processing \citet{Kandel00_book}.
Accordingly, stimulus information is encoded in the timing and frequency
of those spikes \citet{Softky93,Shadlen95,Mainen95_1503,Rieke97,DeRuyterVanSteveninck97,Strong98_197}.
This information transmission is canonically estimated by the mutual
information between the stimulus and the neuronal spiking response
\citet{deRuytervanSteveninck88_379,Rieke97}. In case of discrete
and static stimuli this is done by counting the number of spikes emitted
in a temporal window of a few hundreds of milliseconds following the
stimulation. From the distribution of those spike counts it is then
possible to estimate their empirical entropies and therefore the mutual
information. In many applications, however, input and output are dynamical,
information is transmitted \footnote{From a mathematical point of view, this word is a bit imprecise because
it implies a directionality not reflected in the definition of the
mutual information, which is symmetric in in- and output. Therefore,
``shared'' would be a more neutral way to express this relation.
However, in the practical applications we have in mind, involving
a stimulus as input and a neural activity as output, this directionality
is given by the setup, so we will also use the term ``transmitted''.} over a certain period of time \citet{DeRuyterVanSteveninck97,Strong98_197}
and the static method for estimating mutual information is inappropriate.
In this case, it makes sense to consider the mutual information per
time - the mutual-information rate (MIR \citet{Shannon48_379}) -
or the mutual information per emitted spike \citet{Eckhorn74_191,Strong98_197,Borst02_213,Borst03_23,Koch04_1523,Koch2006}.
Being able to reliably quantify this information is a necessary step
to develop a quantitative understanding of sensory processing. 

In order to achieve the goal of estimating the MIR,  the data-intense
histogram method to determine the entropies of long spiking patterns
has been applied in several works \citet{Strong98_197,Borst02_213,Borst03_23,Koch04_1523,Koch2006}.
Because this approach, also known as direct method, suffers in the
data-limited case, a number of techniques have been developed to regularize
the estimation and correct biases \citet{Treves95_399,Strong98_197}.
These improvements have extended the range of applicability of the
histogram method, which is however still limited to cases with relatively
large datasets. Recently, Mahuas et al. \citet{Mahuas23_024406} have
proposed a complementary approach that avoids histograms developing
MIR as a series in the empirical correlations. As this approach requires
binarized neurons, it can deal only with small time bins, which limits
its applicability.

In this work we introduce the moment-based mutual-information-rate
approximation (Moba-MIRA), a method for estimating MIR from noisy
data  generalizing the work of \citet{Mahuas23_024406}. Our method
grounds on a diagrammatic expansion of the entropies in term of correlations.
Recent results in field theory \citet{Kuehn18_375004,Kuehn23_115001,Kuehn25b_arxiv}
allow us to expand around any non-interacting theory, like Ising spins
as in \citet{Mahuas23_024406}, but also the empirical-spike-count
(integer) distribution, and solving the issue occurring of small time
bins.

We apply our method on data from the retina, a part of the nervous
system that is particularly adapted to being studied by means of information
theory because its input is very well controlled and it is relatively
well accessible for recordings of neural activity. In the retina,
incoming light is first absorbed by photoreceptors, transformed into
an electrical signal, processed by a sequence of neurons and eventually
encoded by the spiking activity of retinal ganglion cells (RGC) \citet{Gollisch10_150}.
 It has been attempted in several studies to estimate the resulting
$\MIR$ by direct methods and Gaussian approximations \citet{Eckhorn74_191,Eckhorn75_7,Koch04_1523,Puchalla05_493,Koch2006,Passaglia04_1217},
yet, as indicated before, the approaches in these works were data
intensive and have therefore a limited range of applicability. 
While we use our technique for the retina as a handy model system,
it is much more widely applicable. One can use it not only for other
neural systems, but virtually any system, not necessarily biological,
whose behavior can be described as the response in form of a stochastic
integer number to some input.

 After briefly discussing the definition of the mutual-information
rate in \secref{Background}, we derive Moba-MIRA in \secref{Results:-method}.
We first test its accuracy, in \subsecref{Test_ArtData}, by applying
it to synthetic data for which the ground truth value of the $\MIR$
can be estimated. In \secref{Results_retinal_data}, we perform an
in-depth analysis of retinal recordings, across stimuli and cell types
and conclude in \secref{Discussion}, giving an outlook on possible
further developments.

\section{Background\label{sec:Background}}

\begin{figure}
\includegraphics[width=1\textwidth]{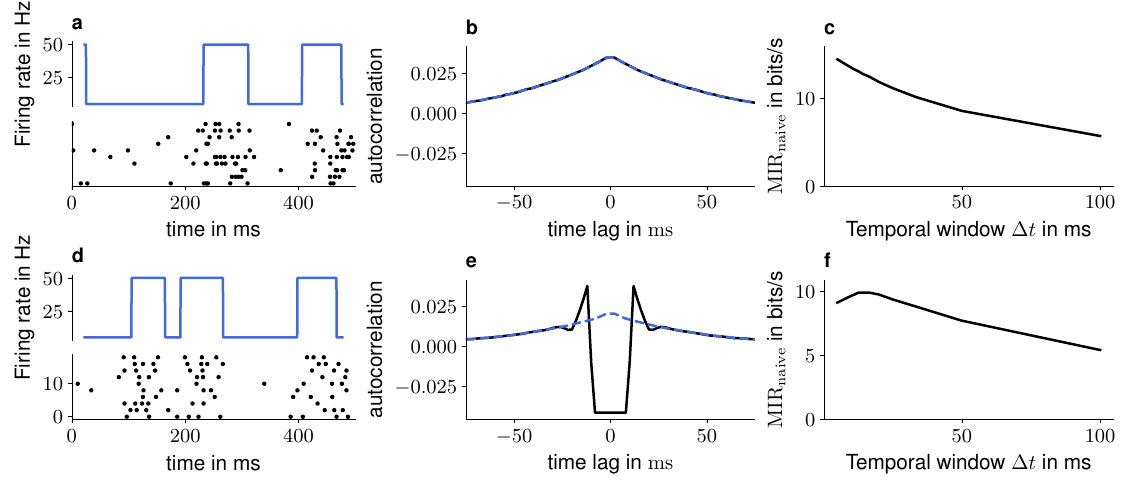}\caption{\textbf{(a)} A neuron emitting spikes according to a rate switching
randomly (Poisson process) between two levels, whereas the spiking
conditional on the rate is either Poissonian itself (also panels b
and c) or \textbf{(d)} features a refractory period (also panels e
and f), a dot indicates a spike. \textbf{(b,e)} Autocorrelations of
the mean activity indicated by the dotted blue lines, autocorrelations
of the spikes by the solid black lines. \textbf{(c,f)} Mutual-information
rates computed lumping activity into one time bin, as in \eqref{Def_MIR_naiv},
as a function of the time-bin size $\Delta t$.\textbf{ }Parameters:
Correlation time of switching of the rate: $T_{\mathrm{switch}}=100\,\protect\ms$,
firing rate $f_{\mathrm{low}}=5\mathrm{Hz}$, $f_{\mathrm{high}}=50\mathrm{Hz}$,
(absolute) refractory period $t_{\mathrm{ref.}}=10\,\mathrm{ms}$,
simulation time $T_{\mathrm{total}}=3\cdot10^{4}\,\protect\ms$, $N_{\mathrm{rep.}}=10^{4}$.\textbf{
}\label{fig:Naiv_definition}}
\end{figure}

How does one compute the rate of transmitted information? A naïve
approach to the problem, that follows from the static case, is to
define MIR as
\begin{equation}
\MIR_{\mathrm{na\"ive}}\coloneqq\frac{\MI\left(\Delta t\right)}{\Delta t},\label{eq:Def_MIR_naiv}
\end{equation}
that is, the mutual information between stimulus and spike count,
binned with a bin size $\Delta t$, and divided by $\Delta t$, where
\begin{equation}
{\cal I}\left(\Delta t\right)=S^{\mathrm{out}}\left(\Delta t\right)-\left\langle S_{\xi}^{\mathrm{in}}\left(\Delta t\right)\right\rangle ,\label{eq:Def_MutualInfo}
\end{equation}
with $S^{\mathrm{out}}\left(\Delta t\right)$ the output (or marginal)
entropy and the input (or conditional) entropy $S_{\xi}^{\mathrm{in}}\left(\Delta t\right)$
conditional on the stimulus. The choice of $\Delta t$ should depend
on the system dynamics, and in particular on the relevant time scale
of the stimulus. Yet, the dynamics of biological systems as the retina
extends over multiple time scales, and cannot be captured by a single
time bin. In order to understand the consequences of this choice,
we consider the toy example of a neuron firing according to an inhomogeneous
Poisson process. Mimicking the effect of a dynamical stimulus, the
neuron's rate randomly switches between a low and high state with
an average frequency of $27\,\mathrm{Hz}$, cf. \figref{Naiv_definition}a.
This yields an exponential decay of the autocorrelation of the neuron,
both for the mean activity over repetitions (peristimulus time histogram,
PSTH) and for its spike count (respectively, blue dashed and black
line in \figref{Naiv_definition}b). Using \eqref{Def_MIR_naiv} to
compute the MIR for this process leads to an estimate monotonously
decaying with the time-bin size (\figref{Naiv_definition}c), showing
that the choice of the bin duration strongly affects the MIR estimation.
For large $\Delta t$ the stimulus dynamics is averaged out, and the
MIR vanish. Upon decreasing $\Delta t$ this effect reduces, with
the estimate for the mutual-information rate monotonously attaining
a limiting value for $\Delta t\rightarrow0$.  However, this is not
a generic behavior, as we figure out by investigating the same system,
but after adding a refractory period to the model neuron, i.e. it
stays silent for $10\,\mathrm{ms}$ after every spike, cf. \figref{Naiv_definition}d.
The other parameters are unchanged. This leads to lower firing rates,
more regular spike trains and spike-count autocorrelations which are
negative and large for small times (\figref{Naiv_definition}e). Refractory
periods decrease the neurons' variability, therefore increasing their
capacity of transmitting information at fixed firing rate \citet{Ferrari18_NeurComp}.
In order for the effect of the refractory period to become noticeable
in the $\MIR$, however, $\Delta t$ has to be large enough (for a
more detailed explanation see \appref{Refractoriness} in the appendix).
Consistently, the naïve MIR is not monotonic anymore and shows a maximum
at around $15\,\mathrm{ms}$ (\figref{Naiv_definition}f). While in
the limit of $\Delta t\rightarrow0$ we avoid stimulus averaging,
we will neglect the positive effect of refractoriness, leading to
an underestimation of the MIR. So, already in the case with only two
time scales, there is no good choice for the length of $\Delta t$
fully accounting for the system dynamics. 

\begin{figure}
\includegraphics[width=0.95\textwidth]{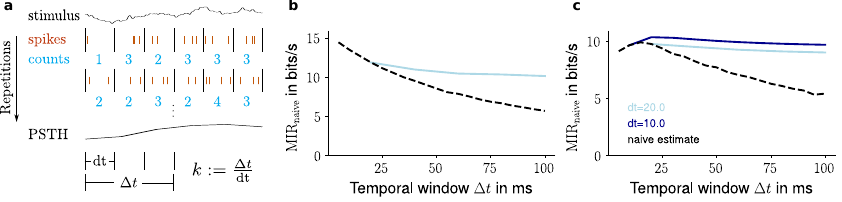}\caption{\textbf{(a)} Sketch of a recording of spiking neurons responding
to a repeated stimulus over a time span $\Delta t$ with the activity
discretised into bins of length $\mathrm{dt}$. \textbf{(b)} estimates
of the $\protect\MIR$ as \eqref{MIR_def_proper}, computed by the
(mixed) Moba-MIRA method, introduced in \secref{Results:-method},
for varying recording time $\Delta t$, along with their naive estimates,
\eqref{Def_MIR_naiv}, as shown in \figref{Naiv_definition}, panel
c. Time-bin size $\protect\dt=20\mathrm{ms}$.\textbf{ (c) }As panel
b, but with refractory period, as in \figref{Naiv_definition}f. Other
parameters as in \figref{Naiv_definition}.\label{fig:Toy_model_proper_definition}}
\end{figure}

In order to solve the problem of multiple time scales, it was proposed
in previous works to compute the MIR as \citet{Shannon48_379,Eckhorn74_191,Borst02_213}:
\begin{equation}
\mathrm{MIR}\coloneqq\lim_{\dt\to0}\underset{\Delta t\rightarrow\infty}{\lim}\frac{{\cal I}\left(\dt,\Delta t\right)}{\Delta t},\label{eq:MIR_def_proper}
\end{equation}
where the mutual information is computed over a large temporal window
$\Delta t$, but after binning the neuron's activity in small consecutive
bins $\dt$, \figref{Toy_model_proper_definition}. Subdividing $\Delta t$
in $k$ time bins, as sketched in \figref{Toy_model_proper_definition}a,
we obtain the estimates of the $\MIR$ for the model introduced in
\figref{Naiv_definition} and plot it in \figref{Toy_model_proper_definition}b,c
as functions of $\Delta t$. We observe that the estimates converge
to constant, non-zero values for $\Delta t\rightarrow\infty$.  The
dependence of the $\MIR$ estimate on $\Delta t$ can be explained
as follows: the estimates of $\MIR$ initially decrease because information
is encoded (slightly) redundantly in the time bins, so that adding
more of them increases the mutual information (slightly) sublinearly.
For large $\Delta t$  the activity in the newly added time bins is
sufficiently distant from the first so that adding more time bins
does not change the estimate of $\MIR$: we reach convergence. For
the process including a refractory period, we have the same behavior
for large $\Delta t$. For small $\Delta t$, we observe an initial
increase with $\Delta t$ of the estimate of the $\MIR$ (panel c),
which is absent without refractoriness. Qualitatively this is the
same behavior as of the naive estimate of the $\MIR$ shown in \figref{Naiv_definition}.
It can be explained by the fact that the activity, which is positively
correlated between adjacent time bins for small $\Delta t$, gets
decorrelated by the refractoriness. More precisely, the refractory
period leads to negative noise (auto-)correlations canceling the positive
stimulus (auto-)correlations. We discuss this effect in more detail
in \appref{Refractoriness}. In the Poisson process without refractory
period (panel b), the spike counts can be very high already for $\dt=10\,\mathrm{ms}$
and therefore, computing the $\MIR$ for a large value for $k$ becomes
infeasible. We therefore have chosen $\dt=20\,\ms$ for panel c. 

The required entropies for the $\MIR$s as shown in \figref{Toy_model_proper_definition}
are those of probability distributions of stochastic paths across
multiple time bins. They are high-dimensional objects because $\Delta t$
has to be large enough to cover the correlation time of the stimulus
(and the response) and $\dt$ has to be small enough to capture the
dynamics of the neural activity. For a limited amount of data, as
is typical  for experiments, these entropies quickly become difficult
or impossible to compute reliably. The direct method to determine
them, namely counting the occurrence of patterns (histogram method
\citet{Strong98_197}) can therefore only work in simple setups, in
particular for single neurons \citet{DeRuyterVanSteveninck97,Borst02_213,Borst03_23,Koch04_1523,Koch2006}.
Even then, the number of time bins is limited (up to eight in the
cited examples)  because the number of possible words grows exponentially
in this quantity. This issue applies in particular to the input entropy
(conditional on the stimulus, cf. \eqref{Def_MutualInfo}), because
it quantifies the variability given a stimulus. It can therefore only
be measured by repeating the same stimulus several times and collecting
statistics over these repetitions. Here, therefore, number of samples
equals that of the repetitions of the same stimulus, so typically
at most about 100 in real data. In contrast, the output entropy marginalizing
over the stimulus is computed over all times and repetitions, therefore
the underlying probability distribution is much better sampled. 

Even though there are sophisticated methods to improve the histogram
method \citet{Strong98_197,Nemenman01,Nemenman11_2013,Hernandez19,Hernandez22,Hernandez23_014101},
it is based on the characterization of a very high-dimensional probability
distribution corresponding to an exponential number of moments. This
is different if the data is Gaussian, which means that it is completely
characterized by its first two moments. In this case there are closed-form
expressions for the corresponding entropies vastly simplifying its
computation \citet{deRuytervanSteveninck88_379,Bialek90_103,Borst99_947,Tostevin10_061917}.
However, this prerequisite is often not met. Another option to regularize
computations for the entropies consists in assuming a model for the
underlying stochastic process, which can be used to parameterize the
corresponding probability distributions. This is the case, for example,
for chemical reaction networks, for which a host of new methods to
compute the $\mathrm{MIR}$ have been developed in recent years \citet{Sinzger23_6822,Moor23_013032,Reinhardt23_041017}.
However, such a model is not available in all situations and even
if it is, it has to be fitted to the data, which can be a non-trivial
- or at least numerically expensive - step on its own.

\section{Results: Moba-MIRA, a robust method to compute the MIR \label{sec:Results:-method}}

\subsection{Maximum-entropy modeling}

 Estimating entropies for real data is difficult because it provides
only a limited number of repetitions, which leads to large biases.
Our suggestion to mitigate this problem is to compute the entropies
of the activities in the single time bins and to approximate the impact
of correlations between them only on a pairwise level. Performing
an expansion in the corresponding pairwise correlations and partly
resumming this series leads to the following approximation for the
entropy:
\begin{equation}
S\approx S_{0}+\frac{1}{2}\left(\ln\left(\det\left(c\right)\right)-\ln\left(\det\left(V\right)\right)\right),\label{eq:Entropy_single_plus_loops}
\end{equation}

where $S_{0}$ is the entropy of all single bins summed up, neglecting
correlations, $c$ is the covariance matrix between bins (the autocovariance
of the neuron under scrutiny across time) and $V$ are the respective
variances, written as the entries of a diagonal matrix. This is what
we call the moment-based mutual-information-rate approximation (Moba-MIRA).
More precisely, we will use two versions of it: in one, we will use
\eqref{Entropy_single_plus_loops} only for the input entropy, while
estimating the output entropy by the histogram method. We will call
this variant the mixed Moba-MIRA, whereas we christen full Moba-MIRA
the variant for which we use \eqref{Entropy_single_plus_loops} for
both types of entropies.

The approximation of \eqref{Entropy_single_plus_loops} makes sense
intuitively: it is nearly the Gaussian approximation, but with the
important difference that we exactly take into account the entropies
of the activitities (not necessarily binary) in single time bins.
In the following we derive \eqref{Entropy_single_plus_loops} in more
grounded way using results from statistical physics and a diagrammatic
expansion and explain how to add additional higher-order corrections.

\subsubsection{Derivation of the Moba-MIRA approximation}

Given a vector $\boldsymbol{n}=\left(n_{1},\dots,n_{k}\right)$ of
spike counts recorded for the duration of $k=\frac{\Delta t}{\dt}$
time bins, depending on some other variable (e.g. a stimulus), which
is identically repeated $N_{\mathrm{rep}}$ times, we want to estimate
the probability distribution $P\left(\boldsymbol{n}\right)$ and the
corresponding entropy $S$. We formalize the separation between statistics
of the single bins, which we treat exactly, and the correlations between
then, which we treat on a pairwise level, by making the following
ansatz: 

\begin{equation}
P\left(\boldsymbol{n}\right)\sim e^{\frac{1}{2}\sum_{t\neq t^{\prime}}n_{t}J_{tt^{\prime}}n_{t}}\prod_{t=1}^{k}e^{-H_{t}\left(n_{t}\right)},\,\boldsymbol{n}\in\mathbb{N}^{k},\label{eq:Def_generalized_Ising}
\end{equation}
where $H_{t}$ is some function and $\left\{ J_{tt^{\prime}}\right\} _{1\leq t<t^{\prime}\leq k}$
is a matrix, which are both determined in order to match the measured
statistics. Concretely, we choose
\begin{equation}
H_{t}\left(n\right)=\sum_{i=1}^{\infty}\lambda_{i}n^{i},\label{eq:Hamiltonian_infinitely_many_coeff}
\end{equation}
which is what results from a maximum-entropy modeling approach \citet{Jaynes03}
with all moments of the single-bin statistics fixed. It is the formal
expression of the informal statement that we treat the single-time-bin
statistics exactly. At first sight, it appears that an infinity of
quantities has to be measured. In practice however, this is not the
case because the number of spikes observed per time bin is of course
finite for finite recordings. Therefore there are at most $kn_{\mathrm{max}}$
quantities to determine for this contribution to the entropy, with
$n_{\mathrm{max}}$ being the maximum number of spikes observed in
one time bin. Together with the $\frac{k\left(k-1\right)}{2}$ correlations,
this yields a number growing only quadratically with $k$. This is
much better than the histogram method with its $\left(n_{\mathrm{max}}+1\right)^{k}$
parameters, which leads to considerable biases when $N_{\mathrm{rep}}=\mathcal{O}\left(10^{2}\right)$
- the relevant regime for experiments. Our method, in contrast, performs
well there, as we will demonstrate later. 

Our approach is the natural generalization of the traditional maximum-entropy
framework employing binary variables \citet{schneidman2006weak},
for which, as well, the single-unit statistics is reproduced (because
fixing the mean already fixes the whole one-parameter distribution).
Choosing binary variables, however, allows to take into account maximally
one spike in every bin, otherwise information is lost. Sticking to
this convention would therefore limit us in the choice of the time-bin
width $\mathrm{dt}$. This would be unfortunate for our purposes,
because we want to study the behavior of our estimate for the $\MIR$
in dependence on arbitrary $\mathrm{dt}$.

One possibility to use \eqref{Def_generalized_Ising} would now be
to fit the parameters $J_{tt^{\prime}}$ and in $H_{t}$ and then
sample from $P$ to compute the entropy. However, we can get around
this step by leveraging recently developed techniques from statistical
field theory \citet{Kuehn18_375004,Kuehn23_115001,Kuehn25b_arxiv},
explained in more detail in app. \ref{app:App_LegendreTrafos-1}.
They allow us to derive the approximation \eqref{Entropy_single_plus_loops}
as a resummation of a class of diagrams in the diagrammatic small-correlation
expansion around the case of uncorrelated time bins

\begin{fmffile}{Gamma_loop_expansion}	 	
	\begin{align} 	       
		S \approx  & 
		S_{0}- \mkern-30mu 
		\parbox{25mm}{ 			
			\begin{fmfgraph*}(75,25) 				
				\fmfpen{0.5thin} 				
				\fmftop{o1,o2,o3,o4,o5} 				
				\fmfbottom{u1,u2,u3,u4,u5} 				
				\fmf{phantom}{u1,v1,o3} 				
				\fmf{plain}{v1,o3} 				
				\fmf{phantom}{o1,v1,u3} 				
				\fmf{plain}{v1,u3} 				
				\fmf{phantom}{u3,v2,o5} 				
				\fmf{plain}{u3,v2} 				
				\fmf{phantom}{o3,v2,u5} 				
				\fmf{plain}{o3,v2} 				
				\fmfv{decor.shape=circle,decor.filled=empty, decor.size=6.5thin}{v1,v2} 			
			\end{fmfgraph*} 			
		} 		
		\mkern-20mu + \mkern-30mu 
		\parbox{25mm}{ 			
			\begin{fmfgraph*}(75,25) 				
				\fmfpen{0.5thin} 				
				\fmftop{o1,o2,o3} 				
				\fmfbottom{u1,u2,u3} 				
				\fmf{phantom,tension=100}{u1,dl,v1,o2} 				
				\fmf{plain}{dul,v1,o2} 				
				\fmf{phantom,tension=100}{u3,dr,v2,o2} 				
				\fmf{plain}{dur,v2,o2} 				
				\fmf{phantom}{u1,dul,u2,dur,u3} 				
				\fmf{plain}{dul,u2,dur} 				
				\fmf{phantom,tension=0.5}{dl,dum,dr} 				
				\fmf{phantom,tension=1}{dum,u2} 				
				\fmfv{decor.shape=circle,decor.filled=empty, decor.size=6.5thin}{v1,v2,u2} 			
				\end{fmfgraph*} 		
			}  
			\mkern-15mu  - \mkern-30mu 
		\parbox{25mm}{ 			
			\begin{fmfgraph*}(75,25) 				
				\fmfpen{0.5thin} 				
				\fmftop{o1,o2,o3} 				
				\fmfbottom{u1,u2,u3} 				
				\fmf{phantom}{u1,dul,dml,v1,o2} 				
				\fmf{plain}{dml,v1,o2} 				
				\fmf{phantom}{u2,v2,dml,dol,o1} 				
				\fmf{plain}{u2,v2,dml} 				
				\fmf{phantom}{u2,v3,dmr,dor,o3} 				
				\fmf{plain}{u2,v3,dmr} 				
				\fmf{phantom}{u3,dur,dmr,v4,o2} 				
				\fmf{plain}{dmr,v4,o2} 				
				\fmfv{decor.shape=circle,decor.filled=empty, decor.size=6.5thin}{v1,v2,v3,v4} 			
				\end{fmfgraph*} 		
			} 		
			\mkern-20mu +\dots,
		\label{eq:Approx_SingleBinEnt_Loops}	 	
	\end{align} 
\end{fmffile}

which, using Feynman-diagrammatic rules \footnote{In statistical physics one sometimes also refers to them as Mayer
diagrams \citet{Mayer77,vasiliev2019functional}. The work of Mayer
\& Goeppert-Mayer is actually a few years older than that of Feynman.} indeed yields \eqref{Entropy_single_plus_loops}. Note that the elements
of the diagrams represent combinations of the measured cumulants -
we do not have to know $J_{tt^{\prime}}$ or $H_{t}$ explicitly.
We will sketch the basics of diagrammatics in appendix \appref{App_LegendreTrafos-1}
and refer to \citet{Kuehn18_375004,Kuehn23_115001,Kuehn25b_arxiv}
for a more detailed description. Considering corrections to \eqref{Approx_SingleBinEnt_Loops}
by taking into account more diagrams can be beneficial in some cases,
but we have found the resummed-loop approximation to be the most robust.

\subsection{Testing the approximation on artificial data\label{subsec:Test_ArtData}}

In order to validate MoBa-MIRA on a biologically plausible example,
we use a generalized linear model (GLM) fitted to generate spike trains
resembling those of retinal ganglion cells, for which we adapt the
setup employed in \citet{Mahuas23_024406} (see there and appendix
\appref{Adapting_Mahuas} for details). As shown in \figref{Comparison_real_artificial},
when properly fitted, the GLM generates interspike-interval distributions
and PSTHs barely distinguishable from real data. At the same time,
when repeating a given stimulus to estimate the input entropies, we
are only limited by the capacities of our computer. This allows us
to compute a numerically exact value for biologically plausible data
as ground truth. 

\begin{figure}
\includegraphics[width=0.95\textwidth]{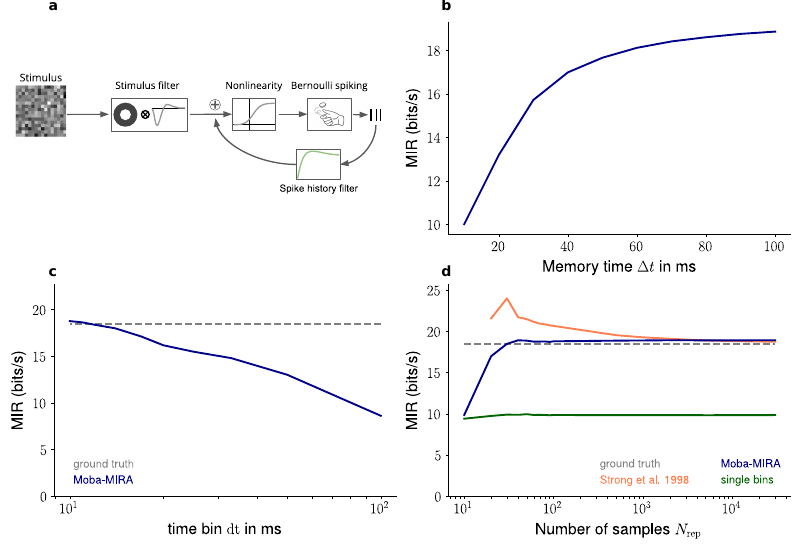}\caption{Test of our method on data from a generalized linear model (GLM).
\textbf{(a)} Scheme of a GLM: firing rates are computed depending
on a visual stimulus, according to which spikes are generated in a
random way (cf. app. \appref{Adapting_Mahuas} for details).  \textbf{(b)}
Estimate of the $\protect\MIR$ in dependence of $\Delta t$ for $\protect\dt=10\,\protect\ms$
and $N_{\mathrm{rep}}=3\cdot10^{4}$. \textbf{(c)} Dependence of the
$\protect\MIR$ on the time-bin size $\protect\dt$ for $\Delta t=100\,\mathrm{ms}$
and $N_{\mathrm{rep}}=3\cdot10^{4}$. \textbf{(d)} Dependence of the
$\protect\MIR$ estimates on the number of repetitions $N_{\mathrm{rep}}$,
$\protect\dt=10\,\mathrm{ms}$, $\Delta t=100\,\mathrm{ms}$. All
results were obtained with mixed Moba-MIRA, they are similar for full
Moba-MIRA, compare \figref{Validation_on_ArtData_Detail}a. For parameters
of the GLM consult app. \appref{Adapting_Mahuas}.\label{fig:Validation_on_ArtData_Overview}}
\end{figure}

In this setup, we can study in detail how the estimate of $\MIR$
changes as a function of $\dt$ and $\Delta t$. As visible from \figref{Validation_on_ArtData_Overview}a,
the estimate has converged at about  $\Delta t=100\,\mathrm{ms}$.
Fixing this value and varying $\dt$, we observe in \figref{Validation_on_ArtData_Overview}b
that we reach convergence at about $\dt=10-15\,\mathrm{ms}$. We therefore
fix $\Delta t=100\,\mathrm{ms}$ and $\dt=10\,\mathrm{ms}$ for \figref{Validation_on_ArtData_Overview}d,
in which we plot the dependence of different estimates on the number
of repetitions. Whereas the direct method shows a clearly visible
bias at $N_{\mathrm{rep}}=\mathcal{O}\left(10^{2}\right)$, Moba-MIRA
is already converged at about $N_{\mathrm{rep}}=50$. This is not
a trivial consequence of negligible correlations between time bins
as the comparison with the result neglecting correlations - that is,
only summing up the entropies of single bins - shows. 

\begin{figure}
\includegraphics[width=0.95\textwidth]{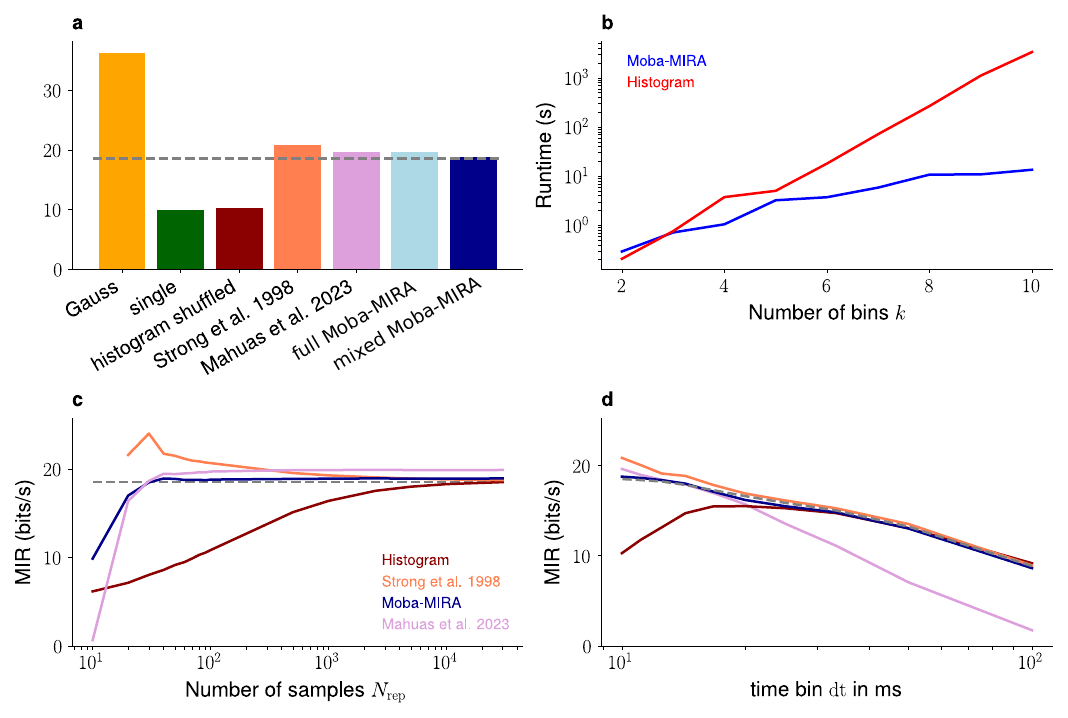}\caption{Benchmark of our method. \textbf{(a)} Estimates of different methods
for $N_{\mathrm{rep}}=80$, ground truth computed with histogram method
with $N_{\mathrm{rep}}=3\cdot10^{4}$. \textbf{(b)} Run time for the
input entropy for the different methods is shown, $N_{\mathrm{rep}}=1000$.
\textbf{(c)} Dependence of the $\protect\MIR$ estimate on $N_{\mathrm{rep}}$,
as panel d of \figref{Validation_on_ArtData_Overview}, but including
the approach from \citet{Mahuas23_024406} and the histogram method
with the same de-biasing procedure applied as for Moba-MIRA (shuffling,
see \secref{Appendix_bias_removal}).   \textbf{(d)} Dependence of
the $\protect\MIR$ estimate on the time bin $\protect\dt$.\label{fig:Validation_on_ArtData_Detail}}
\end{figure}

How does the dependence of $\MIR$ estimates on $\dt$ and $\Delta t$
in the regime of limited data looks like for other methods? In \figref{Validation_on_ArtData_Detail}a,
we demonstrate that assuming the data to be simply Gaussian, which
amounts to setting $S=\ln\left(\det\left(c\right)\right)+\mathrm{const.}$
leads to a clear overestimation of the $\MIR$ (yellow bar). Also,
just neglecting correlations between time bins is not feasible, as
the green bar shows, indicating that actually about half of the information
is captured by the interaction between bins. Characterizing this relation
using the histogram method, assuming a quadratic dependence on $1/N_{R}$
and extrapolating to $1/N_{R}=0$, as suggested by Strong et al. \citet{Strong98_197},
yields much better results (orange bar), but leads still to an overestimate.
However, assuming a probability distribution of the form \eqref{Def_generalized_Ising}
for the probability distribution conditional on the stimulus and applying
the approximation \eqref{Entropy_single_plus_loops} (mixed Moba-MIRA)
yields an excellent fit, provided that we remove the bias by subtracting
the estimates obtained from shuffled data, see \secref{Appendix_bias_removal}
for details. Imposing the form of \eqref{Def_generalized_Ising} for
the output entropy as well (full Moba-MIRA) yields a slightly worse
fit, however, still performs better than the method from \citet{Strong98_197}.

 Why does our approach, \eqref{Entropy_single_plus_loops}, work?
In general pairwise probability distributions, \eqref{Def_generalized_Ising},
might not be appropriate because the data could be also shaped by
higher-order interactions in addition. However, for the input entropy,
the most important part of the covariance is due to the refractory
period of the neurons, as visible in the auto-correlation of the spikes,
cf. \figref{Comparison_real_artificial}. In other words, given
the stimulus, the activity in two time bins is correlated mostly because
a spike in one time bin suppresses a spike in another one \footnote{Note also that this interaction is not directed in time - whenever
there is a spike at some point in time, one knows that there could
not have been a spike both before and after this time}. This effective suppression of spikes in neighbored time bins is
an intrinsically two-point like interaction - at least as long as
the time bin is not considerably shorter than the refractory period.
It therefore does not come as a surprise that this approach works
well for the input entropy. For the output entropy, however, we cannot
make a similar argument and indeed, the pairwise approach clearly
works worse in this case, in particular for smaller time bins.


\section{Results: Application of Moba-MIRA on retinal stimulus response\label{sec:Results_retinal_data}}

\begin{figure}
\includegraphics{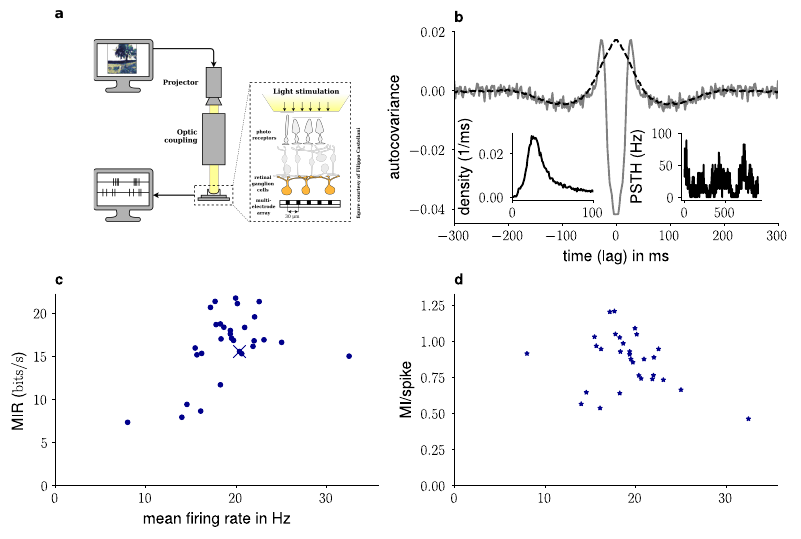}\caption{The $\protect\MIR$ computed with (full) Moba-MIRA from ex-vivo retina
recordings. \textbf{(a)} Sketch of the experimental setup. \textbf{(b)}
Statistical measures of an example neuron - autocorrelation of the
PSTH (gray) and the spikes (black), distribution of the interspike-intervals
in the left inset, the PSTH for an example period in the right inset.
\textbf{ (c)} Scatter plot of the $\protect\MIR$ against the firing
rate. We indicate by a cross the neuron analyzed in more detail in
\figref{MIR_vs_dt_RealData}a. \textbf{(d)} Mutual information per
spike, same data as for panel c. \label{fig:Real_data_MainText}}
\end{figure}

We now apply our method to data recorded in ex-vivo experiments on
rat retinas \citet{Deny17_1964}. To generate this data, the dissected
retinas were stimulated by different light patterns and the activity
of their output layer, containing the ganglion cells, was recorded
by a multi-electrode array (MEA), see \figref{Real_data_MainText}a.
After spike-sorting \citet{Yger18_e34518} ganglion cells were clustered
into functional types, resulting in two large populations of OFF and
ON alpha cells \citet{Deny17_1964}. Firstly, we started working on
the former cells and determined the value of $\dt$ and $\Delta t$
from the autocorrelation of spikes and PSTHs (peristimulus time histogram,
\figref{Real_data_MainText}b): the intrinsic dynamics of the neurons
(refractory period) is in the range of 10-20 milliseconds, whereas
the correlation time due to the stimulus is in the range of hundred
milliseconds. 

 Accordingly, we choose $\dt=10\mathrm{ms}$ and $\Delta t=200\mathrm{ms}$
to estimate the values shown in the panels c and d. A closer inspection
of the dependence of the estimates of the $\MIR$ on these parameters,
as done for the artificial data in \figref{Validation_on_ArtData_Overview}c,
shows that this choice indeed yields a sufficient resolution and a
sufficiently large time window, cf app. \ref{app:RealData_computeMIR}.
Comparing the individual $\MIR$s with the firing rates, we observe
a positive correlation (panel c). This relation, however, is sublinear
so that the mutual information per spike decreases with the firing
rate (panel d), in agreement with \citet[their fig. (3)]{Koch2006}.

\begin{figure}
\includegraphics{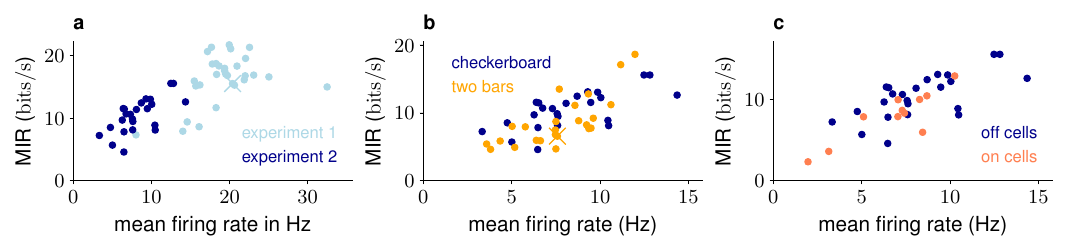}

\caption{Comparison of the relation between firing rates and $\protect\MIR$s
for different modalities. \textbf{(a)} OFF cells recorded during stimulation
with a checkerboard pattern in two different retinal preparations.
\textbf{(b) }Same population of OFF cells subjected to different stimuli.
We indicate by a cross the neuron analyzed in more detail in \figref{MIR_vs_dt_RealData}b.\textbf{
(c)} Populations of ON and OFF cells in the same retinal preparation.\label{fig:Comparison_Scatter_StimExpTypes}}
\end{figure}

We then asked if our findings were consistent across experiments,
stimuli and cell types. Indeed, repeating the same analysis for another
retinal preparation driven by the same stimulus, we observed smaller
firing rates and smaller $\MIR$ values. This is therefore consistent
with the correlation between these two quantities. Also, the same
behavior was observed when comparing the retinal response to two different
visual stimulations - two randomly moving bars and checkerboards (\figref{Comparison_Scatter_StimExpTypes}b).
We observed the same behavior across population of ON and OFF cells
(\figref{Comparison_Scatter_StimExpTypes}c), so that,   again in
agreement to what has been observed in \citet{Koch2006}, there is
no qualitative difference between the two types.

The results we have presented so far were obtained with the ``full''
version of Moba-MIRA, that is, we have used the approximation \eqref{Entropy_single_plus_loops}
for the input and the output entropies. However, in the analysis of
the artificial data, we have seen that mixed Moba-MIRA yields results
visibly closer to the ground truth, cf. \figref{Validation_on_ArtData_Detail}a.
Also, we have found single ON-cells for which full Moba-MIRA seems
to deliver unreliable results, see \secref{Appendix_bias_removal}.
Luckily, the estimates for the $\MIR$ of our real data typically
already converge at values for $\Delta t$ as low as $80\mathrm{ms}$,
see \figref{MIR_vs_dt_RealData}. In this case $k=8$ so that it is
feasible to compute the estimate for $\MIR$ also in the mixed-Moba-MIRA
version. This allows us to compare our results shown in \figref{Real_data_MainText}
and \figref{Comparison_Scatter_StimExpTypes}, obtained by full Moba-MIRA,
by their counterparts obtained by mixed Moba-MIRA, which we show in
\subsecref{Test_ArtData}. We observe that there are no qualitative
differences between these two approaches.

\begin{figure}
\includegraphics[width=1\textwidth]{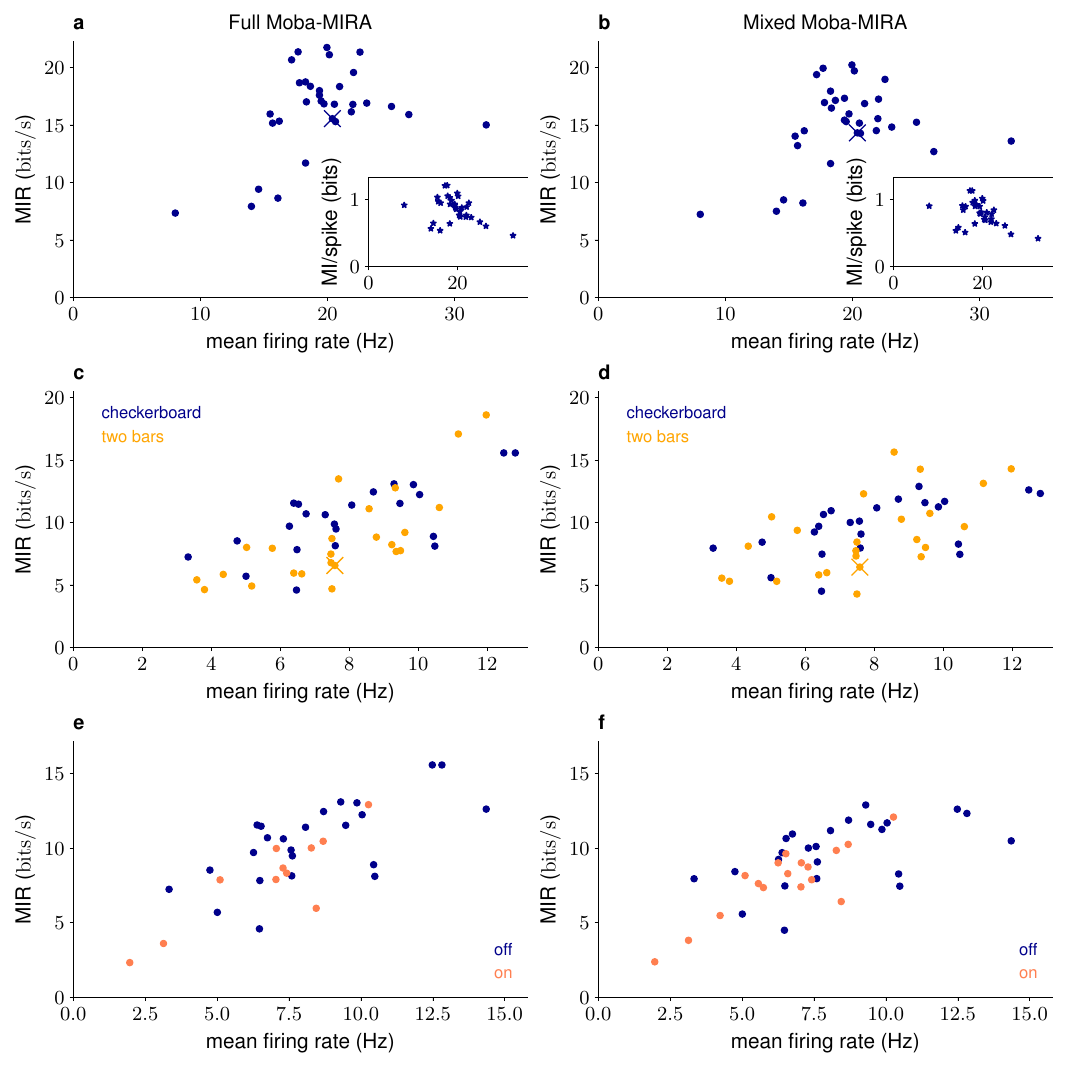}

\caption{\textbf{Left panels)} Full Moba-MIRA and \textbf{Right panels)} mixed
Moba-MIRA. \textbf{(a,b) }Data from \figref{Real_data_MainText},
panels c and panel d (inset). \textbf{(c,d) }Data from \figref{Comparison_Scatter_StimExpTypes},
panel b \textbf{(e,f) }Data from \figref{Comparison_Scatter_StimExpTypes},
panel c.\label{fig:Comparison_full_mixed}}
\end{figure}

\section{Discussion\label{sec:Discussion}}

In this work we address the problem of estimating mutual information
rates for noisy neurons responding to a dynamical stimulus. As observed
in the past \citet{Strong98_197,Koch04_1523}, binning the neuron
activity into large temporal windows fails because the dynamics of
these systems have multiple time-scales, and no matter the bin length,
information at longer or shorter scales is neglected (\figref{Naiv_definition}).
To solve this issue,  short time bins have been used in previous works
\citet{Strong98_197,Borst02_213,Borst03_23,Koch04_1523,Koch2006},
and then long sequences of them were considered, so as to integrate
over all the relevant time scales. These approaches are data hungry,
and therefore very sensitive to undersampling. In contrast, our moment-based
mutual-information-rate approximation (Moba-MIRA) is a method based
on maximum-entropy assumptions relying on the estimate of the single-bin
entropies and low-order statistics, which require much less measurements.
As opposed to related approaches \citet{Tkacik09_arxiv,Mahuas23_024406},
it is not limited to a binary representation of neural activity, therefore
allowing for larger time bins and consequently achieves a compromise
between resolution and the feasibility of computations. Based on a
diagrammatic expansion and a resummation, it allows to robustly estimate
entropies and the mutual-information rate.

The purpose of Moba-MIRA is to estimate conditional and marginal entropy
of the spiking activity. In our approach we assume a pairwise maximum-entropy
distribution of integer neurons and then determine the corresponding
entropy by a small-correlation expansion, similar to the expansion
for the Ising model of \citet{Sessak09}. To derive this approximation,
our framework benefits from recent developments in the field-theory
 for non-Gaussian theories \citet{Kuehn18_375004,Kuehn23_115001}
that allow for using Feynman diagrams to compute corrections, potentially
at all orders. We applied Moba-MIRA to synthetic datasets for which
the ground truth is known, and benchmark it against previous proposed
approaches. Moba-MIRA outperforms them, especially in the data-limited
regime (\figref{Validation_on_ArtData_Overview}d). 

Lastly we applied Moba-MIRA on rat retinal recordings, proving its
capabilities in practical applications. We estimated the mutual information
rate for rat retinal ganglion cells in response to checkerboard and
randomly-moving bars movies. For one experiment for the checkerboard
stimulus, we obtained rates between about $5\,\mathrm{bits}/s$ and
$20\,\mathrm{bits}/s$, corresponding to $0.5\,\mathrm{bits}/\mathrm{spike}$
and $2.2\,\mathrm{bits}/\mathrm{spike}$ (mean equal to $1.1\pm0.4\mathrm{bits}/\mathrm{spike}$,
$n=30$ cells, compare \figref{Comparison_Scatter_StimExpTypes}c).
Our estimates are in the range of previous results from the literature
($2.0\pm0.7\,\mathrm{bits/\mathrm{spike}}$ and $2.1\pm0.6\,\mathrm{bits/\mathrm{spike}}$
for brisk and sluggish cells in the rabbit, respectively \citet{Koch04_1523},
similar range for other cell types \citet{Koch2006}), perhaps slightly
lower. This deviation is mostly because our firing rates are a bit
higher on average, but could also, to a smaller part, be due to the
circumstance that the method by \citet{Strong98_197} tends to overestimate
the $\MIR$, compare also \figref{Validation_on_ArtData_Overview}
and figure A3.

Estimating mutual information rates is a relevant challenge in computational
neuroscience of sensory systems. Previous methods are based on data-intensive
histogram methods \citet{Borst02_213,Borst03_23,Koch04_1523,Koch2006},
which have then been refined with additional extrapolation techniques
\citet{Treves95_399,Strong98_197}. In order to compute information
rates, we followed a different approach, and developed an approximation
scheme that requires only the empirical estimation of several correlation
matrices and that of the single-bin entropies, without needing the
full probability distribution. 

With our approach we extend and generalize \citet{Mahuas23_024406}.
First we consider integer spike counts, instead of binary, allowing
for longer time bins  and therefore less statistics to fit, without
the need for clipping. Additionally, our theoretical scheme allows
for using Feynman rules to compute corrections at potentially all
orders. Note also that due to employing these approximations, we can
compute the entropies directly from easy-to-measure quantities like
covariances, without the need to fit a statistical model or to even
define one. This fitting would be a step with numerically non-negligible
costs, requiring to, e.g., iterate over several rounds of Monte-Carlo
simulations to fix the correct values of the couplings, which we avoid.

Employing Moba-MIRA, we assume that the spike counts over multiple
consecutive time bins follow a pairwise maximum-entropy distribution
\citet{Jaynes03,schneidman2006weak}. These distributions have been
proven effective in modeling of the activity of neural populations
both for marginal \citet{schneidman2006weak,Ferrari17_042321} and
conditional \citet{Ferrari18_PRE,Tkacik09_arxiv} distributions.
As explained before, it is theoretically sound that a pairwise model
works well for the probability distribution conditional on the input
because its correlation structure is mostly determined by the refractory
period after each spike of a neuron. Indeed, if the time-bin size
is of the order of the refractory period, a change in the statistics
of one time bin influences the statistics in the neighbouring bins,
which is an intrinsically pairwise (even local) interaction. It is
also symmetric because the occurrence of a spike in a certain time
bin makes it equally less probable that another spike will occur after
that and that a spike has occurred before. 

Also, the pairwise model \eqref{Def_generalized_Ising} provides even
more potential to reduce numerical costs: By using the alternative
definition 
\begin{equation}
\mathrm{MIR}_{\mathrm{diff}}\coloneqq\underset{\Delta t\rightarrow\infty}{\lim}\lim_{\dt\to0}\frac{{\cal I}\left(\dt,\Delta t+\dt\right)-{\cal I}\left(\dt,\Delta t\right)}{\dt}\label{eq:MIR_diff_Def}
\end{equation}
for the mutual-information rate, one can reduce the numerical cost
from being quadratic in $k$ to linear. Indeed, with the definition
\eqref{MIR_def_proper}, one has to compute the correlations of $\frac{k\left(k-1\right)}{2}$
pairs of bins, assembled from the $k$ time bins in the time window
$\Delta t$. With the definition \eqref{MIR_diff_Def}, however, one
only has to compute the correlation of the activity in the most advanced
time bin, corresponding to the interval $\left(\Delta t,\Delta t+\dt\right]$,
with the $k$ bins inside $\left[0,\Delta t\right]$, therefore only
$k$ correlations. As noted before \citet{Eckhorn74_191}, both definitions
for the MIR are equivalent when the correlations between time bins
decay to $0$. In practice, the achievable $\Delta t$ might not always
be large enough to make both expressions agree, but even then, \eqref{MIR_diff_Def}
is an appealing alternative because it is closer to a differential
definition of a rate or flux. 

Still, however useful the hypothesis that pairwise couplings are
sufficient to describe the interactions across time bins might be
for computations, it does not hold true in general for both distributions,
as a mixture of pairwise MaxEnt distributions does not belong to the
model family itself. To reduce the possible impact of this uncontrolled
assumption, we have proposed a variant of Moba-MIRA (mixed Moba-MIRA),
for which we approximate only the conditional entropies, while performing
extensive histogram count for computing the marginal entropy, where
all the available data points can be used for one entropy estimate.
Even if this comes with additional computational costs, we observed
a neat improvement on the overall performance.

While our theoretical framework allows for computing corrections of
higher order in the pairwise correlations, we did not observe an improvement
of the performance in this case. A possible explanation is that by
assuming a pairwise distribution, we are neglecting higher-order correlations,
and these might have a larger impact than higher order terms in the
expansion in pairwise correlations. Quantifying the relative impact
of all different terms is difficult and would require an extension
of our framework. In principle, however, our expansion around non-Gaussian,
integer neurons allows for including higher-order correlations, and
we will generalise Moba-MIRA to include them in the future. As indicated,
this will be particularly interesting for the output entropy, for
which the pairwise approximation is fair, but not optimal and actually
sometimes fails qualitatively. While we can deal with this problem
by employing the histogram method for the output entropy (that is,
use mixed Moba-MIRA), a faster method for these cases is desirable.

In this work we applied Moba-MIRA to estimate mutual-information rates
of individual neurons. Our method can however be extended to account
for populations by modeling the correlation between different neurons
at different times, going beyond the study of zero-time-lag correlations
and their impact on information \citet{MorenoBote14_1410,AzeredoDaSilveira21_403,Mahuas23_024406,Mahuas25_033012}.
Even if undersampling might be an issue there, we expect Moba-MIRA
to be very useful, as methods based on histogram approaches would
require an even larger amount of data, often beyond existing experimental
datasets. With Moba-MIRA, however, only the estimation of correlations
matrices is required, which reduces the necessary dataset size. We
thus expect that reliable estimates can be given at least for pairs
of neurons, analyzed with a temporal resolution comparable to that
employed in this study.  Currently, extending Moba-MIRA in that direction
can be hindered by the lack of ground truth estimation for large populations,
and because of this we leave it for future developments.

\begin{acknowledgments}
We thank Kyle Bojanek and Olivier Marre and for insightful discussions
and Filippo Castellani for providing the sketch of the experimental
setup in \figref{Real_data_MainText}.

We acknowledge funding by ANR-21-CE37-0024 NatNetNoise, by IHU FOReSIGHT
(ANR-18-IAHU-01) and by Sorbonne Center for Artificial Intelligence-Sorbonne
University IDEX SUPER 11-IDEX0004. This work has been done within
the framework of the PostGenAI@Paris project and it has benefitted
from financial support by the Agence Nationale de la Recherche (ANR)
with the reference ANR-23-IACL-0007. Our lab is part of the DIM C-BRAINS,
funded by the Conseil Régional d’Ile-de-France. 

\appendix
\appendixfigures

\section{Technical details and additional analysis}

\subsection{Computing entropies by a diagrammatic small-correlation expansion
around a theory with given statistics\label{app:App_LegendreTrafos-1}}

For the statistics of a spike train discretized into $k$ different
bins, in each of which there can be up to $n_{\mathrm{max}}$ spikes,
fully characterizing the statistics means assigning a probability
to each of the $\left(n_{\mathrm{max}}+1\right)^{k}$ states (sometimes
called words \citet{Strong98_197}). For big $n_{\mathrm{max}}$ and,
in particular, big $k$, the number of states becomes very large,
which prohibits a reliable estimation given limited data. 

Our way out we suggest in this manuscript is to use less demanding
statistical measures, like the covariance between the activities of
different bins to estimate the entropy. We therefore make the ansatz 
\begin{equation}
P\left(\boldsymbol{n}\right)=\frac{1}{{\cal Z}}e^{\frac{1}{2}\sum_{i\neq j}n_{i}J_{ij}n_{j}}\prod_{i=1}^{N}e^{-H_{i}\left(n_{i}\right)},\label{eq:Basic_form_P}
\end{equation}
where $H_{i}$ is some function to be inferred and 
\[
{\cal Z}=\sum_{\boldsymbol{n}}e^{\frac{1}{2}\sum_{i\neq j}n_{i}J_{ij}n_{j}}\prod_{i=1}^{N}e^{-H_{i}\left(n_{i}\right)}
\]
is the partition function, for the probability distribution of $\boldsymbol{n}$.
The log-likelihood of this distribution is given by 
\[
{\cal L}=\frac{1}{2}\sum_{i\neq j}J_{ij}\left\langle n_{i}n_{j}\right\rangle _{P}-\sum_{i=1}^{N}\left\langle H_{i}\left(n_{i}\right)\right\rangle _{P}-\ln\left({\cal Z}\right),
\]
where by $\left\langle \dots\right\rangle _{P}$ we denote the average
over the empirical distribution. The function $H_{i}$ is arbitrary
in our technical framework - one might choose, e.g., $H_{i}\left(n\right)=-\ln\left(n!\right)-e^{h_{i}}$
in case one would like to expand around a Poissonian theory. We, instead,
implement a maximum-entropy approach choosing it as power series according
to \eqref{Hamiltonian_infinitely_many_coeff}. With this choice, we
force our statistical model to reproduce all empirically measured
cumulants for single bins. This is ensured by choosing the interaction
matrix $J$ and the (infinitely many) parameters of $\boldsymbol{H}$,
$\lambda_{1},\lambda_{2},\dots$ accordingly. Because $\ln\left({\cal Z}\right)$
is the cumulant-generating function, we can express this condition
as 
\[
{\cal L}_{\mathrm{max}}=\sup_{J,\lambda_{1},\lambda_{2},\dots}\left(\frac{1}{2}\sum_{i\neq j}J_{ij}\left\langle n_{i}n_{j}\right\rangle _{P}-\sum_{i=1}^{N}\sum_{\alpha=1}^{\infty}\lambda_{\alpha}\left\langle n_{i}^{\alpha}\right\rangle _{P}-\ln\left({\cal Z}\right)\right),
\]
where $\left\langle \dots\right\rangle _{P}$ is the empirical average.
Once the parameters are fixed such that the measured statistics are
reproduced by the model, we can also write
\[
{\cal L}_{\mathrm{max}}=\sup_{J,\lambda_{1},\lambda_{2},\dots}\left(\frac{1}{2}\sum_{i\neq j}J_{ij}\left\langle n_{i}n_{j}\right\rangle -\sum_{i=1}^{N}\sum_{\alpha=1}^{\infty}\lambda_{\alpha}\left\langle n_{i}^{\alpha}\right\rangle -\ln\left({\cal Z}\right)\right)=-S,
\]
where we denote by $\left\langle \dots\right\rangle $ the average
with respect to the model. In words: the negative maximum log-likelihood
equals the entropy of the statistical model and also the free energy
at fixed covariances and fixed single-neuron statistics.

To compute it in practice, we perform an expansion in small covariances.
We will use Feynman diagrams for it, which simplify this endeavor
because the corresponding rules incorporate the structure of the terms
in the series in a compact and elegant way, which comes about by the
fact that the free energy is a Legendre transform \citet{Kuehn23_115001,Kuehn25b_arxiv}.
Also, they allow to identify contributions to the series according
to certain topologies of the diagrams, which can partly be resummed.
For our work, we employ the resummation of all loops, as in

\begin{fmffile}{Loop_expansion}	 	
	\begin{align} 	       
		- \mkern-30mu 
		\parbox{25mm}{ 			
			\begin{fmfgraph*}(75,25) 				
				\fmfpen{0.5thin} 				
				\fmftop{o1,o2,o3,o4,o5} 				
				\fmfbottom{u1,u2,u3,u4,u5} 				
				\fmf{phantom}{u1,v1,o3} 				
				\fmf{plain}{v1,o3} 				
				\fmf{phantom}{o1,v1,u3} 				
				\fmf{plain}{v1,u3} 				
				\fmf{phantom}{u3,v2,o5} 				
				\fmf{plain}{u3,v2} 				
				\fmf{phantom}{o3,v2,u5} 				
				\fmf{plain}{o3,v2} 				
				\fmfv{decor.shape=circle,decor.filled=empty, decor.size=6.5thin}{v1,v2} 			
			\end{fmfgraph*} 			
		} 		
		\mkern-20mu + \mkern-30mu 
		\parbox{25mm}{ 			
			\begin{fmfgraph*}(75,25) 				
				\fmfpen{0.5thin} 				
				\fmftop{o1,o2,o3} 				
				\fmfbottom{u1,u2,u3} 				
				\fmf{phantom,tension=100}{u1,dl,v1,o2} 				
				\fmf{plain}{dul,v1,o2} 				
				\fmf{phantom,tension=100}{u3,dr,v2,o2} 				
				\fmf{plain}{dur,v2,o2} 				
				\fmf{phantom}{u1,dul,u2,dur,u3} 				
				\fmf{plain}{dul,u2,dur} 				
				\fmf{phantom,tension=0.5}{dl,dum,dr} 				
				\fmf{phantom,tension=1}{dum,u2} 				
				\fmfv{decor.shape=circle,decor.filled=empty, decor.size=6.5thin}{v1,v2,u2} 			
				\end{fmfgraph*} 		
			}  
			\mkern-15mu  - \mkern-30mu 
		\parbox{25mm}{ 			
			\begin{fmfgraph*}(75,25) 				
				\fmfpen{0.5thin} 				
				\fmftop{o1,o2,o3} 				
				\fmfbottom{u1,u2,u3} 				
				\fmf{phantom}{u1,dul,dml,v1,o2} 				
				\fmf{plain}{dml,v1,o2} 				
				\fmf{phantom}{u2,v2,dml,dol,o1} 				
				\fmf{plain}{u2,v2,dml} 				
				\fmf{phantom}{u2,v3,dmr,dor,o3} 				
				\fmf{plain}{u2,v3,dmr} 				
				\fmf{phantom}{u3,dur,dmr,v4,o2} 				
				\fmf{plain}{dmr,v4,o2} 				
				\fmfv{decor.shape=circle,decor.filled=empty, decor.size=6.5thin}{v1,v2,v3,v4} 			
				\end{fmfgraph*} 		
			} 		
			\mkern-20mu +\dots 		 	
		= \mathrm{tr}\left( \sum_{n}^{\infty} \frac{\left(-1\right)^{n}}{2n} \left(\frac{c}{\boldsymbol{V}^{\mathrm{T}}\boldsymbol{V}} V\right)^{n}\right)  
		= \frac{1}{2}\left(\ln\left( \det\left(c\right)\right) - \ln\left(\det\left(V\right)\right)\right),
		\label{eq:Resummation_loops_pure}	 	
	\end{align} 
\end{fmffile}

where we call the vector of variances $\boldsymbol{V}$, understand
$\frac{c}{\boldsymbol{V}^{\mathrm{T}}\boldsymbol{V}}$ as element-wise
product and denote the diagonal matrix filled with the entries of
$\boldsymbol{V}$ as $V$. The infinite sum is represented by Feynman
diagrams, composed of the elements

\begin{fmffile}{Collection_diagram_elements}	
	\begin{eqnarray}
		\parbox[35mm]{25mm}{
			\begin{fmfgraph*}(25,25)
				\fmfpen{0.5thin}
				\fmfleft{l1}
				\fmfright{r1}
				\fmf{plain}{l1,c1,r1}
				\fmfv{decor.shape=circle,decor.filled=empty, decor.size=6.5thin}{c1}
				\fmfv{label=$i$, label.angle=-180, label.dist=4.5pt}{l1}
				\fmfv{label=$i$, label.angle=0, label.dist=2.pt}{r1}
			\end{fmfgraph*}
		}  \mkern-55mu = V_{i}, 
		\mkern 120mu
		\parbox{25mm}{
			\begin{fmfgraph*}(25,25)
				\fmfpen{0.5thin}
				\fmftop{o1,o2,o3}
				\fmfbottom{u1,u2,u3}
				\fmf{plain}{u1,o2}
				\fmf{plain}{u3,o2}
				\fmfv{label=$i$, label.angle=-90, label.dist=4.5pt}{u1}
				\fmfv{label=$j$, label.angle=-90, label.dist=4.5pt}{u3}
			\end{fmfgraph*}
		} \mkern-60mu = \frac{c_{ij}}{V_{i}V_{j}}.\label{eq:Def_elements_diagrams}\\ 
	\end{eqnarray}
\end{fmffile}

Every diagram is evaluated by multiplying the mathematical objects
its components represent, summing over all indices and including a
factor, which depends on the number of vertices (edges) and the symmetry
of the diagram. For a description of the precise Feynman rules, we
refer to \citet{Kuehn18_375004,Kuehn23_115001,Kuehn25b_arxiv}. Note
that it is possible to write the resummation in this compact form,
using matrix operations, because we perform the Legendre transform
not only with respect to the covariances, the off-diagonal part of
the covariance matrix, as in \citet{Kuehn23_115001}, but also with
respect to the variances, its diagonal part (and all other single-bin
cumlants). As explained in detail in \citet[sec. 2.2.3]{Kuehn25b_arxiv},
this lifts restrictions in the sums of the perturbation expansion,
which would prevent writing them as simple matrix multiplications.

The rule to determine the estimation of the entropy for the whole
system is therefore quite straight-forward: Compute the entropies
of the single bins, sum them up, then add correcting terms, coming
about by the correlations between the bins and expressed by diagrams.
By construction, the cumulants in the small-correlation expansion
translated from diagrams are then the empirical ones - because the
unperturbed theory is the maximum-entropy single-bin model reproducing
the single-bin statistics.

\subsection{The effect of a refractory period on estimates of the mutual-information
rate\label{app:Refractoriness}}

We will demonstrate in this section that for spiking activity with
a (hard) refractory period $t_{\mathrm{ref}}$, the estimate for the
$\MIR$ initially grows with $\Delta t$, as visible in \figref{Toy_model_proper_definition}c.

Assume that $\Delta t<t_{\mathrm{ref}}$, so that we only have to
consider a binary representation of the activity and consider $k=2$,
so $\dt=\Delta t/2<t_{\mathrm{ref}}/2$. In this two bins, the firing
rate can attain two different values, which we call $\lambda_{1}$
and $\lambda_{2}$. We assume that the averages of both rates are
identical. We can then explicitly compute the $\MIR$ based on the
activities $n_{1}$ and $n_{2}$ of the corresponding time bins. First,
we obtain for the input entropy
\begin{align*}
S_{\mathrm{in}}\left(\dt,\dt\right)= & -\left(1-\left(\lambda_{1}+\lambda_{2}\right)\dt\right)\ln\left(1-\left(\lambda_{1}+\lambda_{2}\right)\dt\right)\\
 & -\dt\left(\lambda_{1}\ln\left(\lambda_{1}\dt\right)+\lambda_{2}\ln\left(\lambda_{2}\dt\right)\right)
\end{align*}
and for the output entropy
\begin{align*}
S_{\mathrm{out}}\left(\dt,\dt\right)= & -\left(1-\left\langle \lambda_{1}+\lambda_{2}\right\rangle _{\boldsymbol{\lambda}}\dt\right)\ln\left(1-\left\langle \lambda_{1}+\lambda_{2}\right\rangle _{\boldsymbol{\lambda}}\dt\right)\\
 & -\dt\left(\left\langle \lambda_{1}\right\rangle _{\boldsymbol{\lambda}}\ln\left(\left\langle \lambda_{1}\right\rangle _{\boldsymbol{\lambda}}\dt\right)+\left\langle \lambda_{2}\right\rangle _{\boldsymbol{\lambda}}\ln\left(\left\langle \lambda_{2}\right\rangle _{\boldsymbol{\lambda}}\dt\right)\right),
\end{align*}
which then yields for the mutual information
\begin{align}
{\cal I}\left(\dt,\dt\right)= & -\left(1-\left\langle \lambda_{1}+\lambda_{2}\right\rangle _{\boldsymbol{\lambda}}\dt\right)\ln\left(1-\left\langle \lambda_{1}+\lambda_{2}\right\rangle _{\boldsymbol{\lambda}}\dt\right)\nonumber \\
 & +\left\langle \left(1-\left(\lambda_{1}+\lambda_{2}\right)\dt\right)\ln\left(1-\left(\lambda_{1}+\lambda_{2}\right)\dt\right)\right\rangle _{\boldsymbol{\lambda}}\nonumber \\
 & -\dt\left(\left\langle \lambda_{1}\right\rangle _{\boldsymbol{\lambda}}\ln\left(\left\langle \lambda_{1}\right\rangle _{\boldsymbol{\lambda}}\dt\right)+\left\langle \lambda_{2}\right\rangle _{\boldsymbol{\lambda}}\ln\left(\left\langle \lambda_{2}\right\rangle _{\boldsymbol{\lambda}}\dt\right)\right)\nonumber \\
 & +\dt\left(\left\langle \lambda_{1}\ln\left(\lambda_{1}\dt\right)\right\rangle _{\boldsymbol{\lambda}}+\left\langle \lambda_{2}\ln\left(\lambda_{2}\dt\right)\right\rangle _{\boldsymbol{\lambda}}\right).\label{eq:MutualInfo_refrac_TwoBins_raw}
\end{align}
We expand the first two lines in $\dt$ to obtain 
\begin{align*}
 & -\left(1-\left\langle \lambda_{1}+\lambda_{2}\right\rangle _{\boldsymbol{\lambda}}\dt\right)\ln\left(1-\left\langle \lambda_{1}+\lambda_{2}\right\rangle _{\boldsymbol{\lambda}}\dt\right)\\
 & +\left\langle \left(1-\left(\lambda_{1}+\lambda_{2}\right)\dt\right)\ln\left(1-\left(\lambda_{1}+\lambda_{2}\right)\dt\right)\right\rangle _{\boldsymbol{\lambda}}\\
= & -\left(1-\left\langle \lambda_{1}+\lambda_{2}\right\rangle _{\boldsymbol{\lambda}}\dt\right)\left(-\left\langle \lambda_{1}+\lambda_{2}\right\rangle _{\boldsymbol{\lambda}}\dt-\frac{1}{2}\left(\left\langle \lambda_{1}+\lambda_{2}\right\rangle _{\boldsymbol{\lambda}}\dt\right)^{2}\right)\\
 & +\left\langle \left(1-\left(\lambda_{1}+\lambda_{2}\right)\dt\right)\left(-\left(\lambda_{1}+\lambda_{2}\right)\dt-\frac{1}{2}\left(\left(\lambda_{1}+\lambda_{2}\right)\dt\right)^{2}\right)\right\rangle _{\boldsymbol{\lambda}}+\mathcal{O}\left(\dt^{3}\right)\\
= & -\frac{1}{2}\left(\left\langle \lambda_{1}+\lambda_{2}\right\rangle _{\boldsymbol{\lambda}}\dt\right)^{2}+\frac{1}{2}\left\langle \left(\left(\lambda_{1}+\lambda_{2}\right)\dt\right)^{2}\right\rangle _{\boldsymbol{\lambda}}+\mathcal{O}\left(\dt^{3}\right)\\
= & \frac{1}{2}\dt^{2}\left\llangle \left(\lambda_{1}+\lambda_{2}\right)^{2}\right\rrangle _{\boldsymbol{\lambda}}+\mathcal{O}\left(\dt^{3}\right),
\end{align*}
where $\left\llangle \dots\right\rrangle $ indicates the cumulant
of $\dots$. Because we assume the statistics of the firing rate to
be stationary, the last two lines of \eqref{MutualInfo_refrac_TwoBins_raw}
simplify to
\begin{align}
 & 2\dt\left(\left\langle \lambda\ln\left(\lambda\right)\right\rangle _{\lambda}+\ln\left(\dt\right)\left\langle \lambda\right\rangle _{\lambda}-\left\langle \lambda\right\rangle _{\lambda}\ln\left(\left\langle \lambda\right\rangle _{\lambda}\right)-\ln\left(\dt\right)\left\langle \lambda\right\rangle _{\lambda}\right)\nonumber \\
= & 2\dt\left(\left\langle \lambda\ln\left(\lambda\right)\right\rangle _{\lambda}-\left\langle \lambda\right\rangle _{\lambda}\ln\left(\left\langle \lambda\right\rangle _{\lambda}\right)\right).\label{eq:MutualInfo_refrac_two_bins_linear_dt_simplified}
\end{align}
So, in total, we obtain obtain for the mutual information of the pair
of time bins
\[
{\cal I}\left(\left(\dt,\dt\right)\right)=2\dt\left(\left\langle \lambda\ln\left(\lambda\right)\right\rangle _{\lambda}-\left\langle \lambda\ln\left(\lambda\right)\right\rangle _{\lambda}\right)+\frac{1}{2}\dt^{2}\left\llangle \left(\lambda_{1}+\lambda_{2}\right)^{2}\right\rrangle +\mathcal{O}\left(\dt^{3}\right).
\]
Similarly, we obtain for the mutual information of the single bins
\[
{\cal I}\left(\left(\dt\right)\right)=\dt\left(\left\langle \lambda\ln\left(\lambda\right)\right\rangle _{\lambda}-\left\langle \lambda\ln\left(\lambda\right)\right\rangle _{\lambda}\right)+\frac{1}{2}\dt^{2}\left\llangle \lambda^{2}\right\rrangle _{\lambda}+\mathcal{O}\left(\dt^{3}\right).
\]
Therefore, the estimate of the $\MIR$ are, respectively
\begin{align*}
\MIR\left(\dt,\dt\right) & =\frac{{\cal I}\left(\left(\dt,\dt\right)\right)}{2\dt}=\left(\left\langle \lambda\ln\left(\lambda\right)\right\rangle _{\lambda}-\left\langle \lambda\ln\left(\lambda\right)\right\rangle _{\lambda}\right)+\frac{1}{4}\dt\left\llangle \left(\lambda_{1}+\lambda_{2}\right)^{2}\right\rrangle +\mathcal{O}\left(\dt^{2}\right)\\
\MIR\left(\dt\right) & =\frac{{\cal I}\left(\left(\dt\right)\right)}{\dt}=\left(\left\langle \lambda\ln\left(\lambda\right)\right\rangle _{\lambda}-\left\langle \lambda\ln\left(\lambda\right)\right\rangle _{\lambda}\right)+\frac{1}{2}\dt\left\llangle \lambda^{2}\right\rrangle _{\lambda}+\mathcal{O}\left(\dt^{2}\right),
\end{align*}
the difference being
\begin{equation}
\MIR\left(\dt,\dt\right)-\MIR\left(\dt\right)=\frac{1}{2}\dt\left\llangle \lambda_{1}\lambda_{2}\right\rrangle ,\label{eq:Increase_MIR_double_bins}
\end{equation}
which corresponds to the difference between the $\MIR$ estimates
for $k=2$ and $k=1$ (or $\Delta t=2\dt$ and $\Delta t=\dt$ in
\figref{Toy_model_proper_definition}). If the rates in two consecutive
time bins are positively correlated, computing the $\MIR$ for both
of the bins at once increases the estimate, otherwise it decreases
it. This observation reminds of what is known as sign rule in the
study of noise correlations in populations of neurons \citet{Hu14}:
if noise and stimulus correlations have opposite sign, the mutual
information increases compared to the case without noise correlations.
Indeed, refractoriness leads to negative noise correlations and positive
rate correlations leads to an increase of the $\MIR$, whereas negative
rate correlations lead to a decrease.

\subsubsection{Increasing the time bin}

We can also lump together the activity of the two time bins into a
larger one, of size $2\dt$, instead of two consecutive time bins
of size $\dt$, as in \figref{Naiv_definition}. This changes the
above computation a bit. We then have
\begin{align}
{\cal I}\left(2\dt\right)= & -\left(1-\left\langle \lambda_{1}+\lambda_{2}\right\rangle _{\boldsymbol{\lambda}}\dt\right)\ln\left(1-\left\langle \lambda_{1}+\lambda_{2}\right\rangle _{\boldsymbol{\lambda}}\dt\right)\nonumber \\
 & +\left\langle \left(1-\left(\lambda_{1}+\lambda_{2}\right)\dt\right)\ln\left(1-\left(\lambda_{1}+\lambda_{2}\right)\dt\right)\right\rangle _{\boldsymbol{\lambda}}\nonumber \\
 & -\dt\left\langle \lambda_{1}+\lambda_{2}\right\rangle _{\boldsymbol{\lambda}}\ln\left(\left\langle \lambda_{1}+\lambda_{2}\right\rangle _{\boldsymbol{\lambda}}\dt\right)\nonumber \\
 & +\dt\left\langle \left(\lambda_{1}+\lambda_{2}\right)\ln\left(\left(\lambda_{1}+\lambda_{2}\right)\dt\right)\right\rangle _{\boldsymbol{\lambda}}.\label{eq:MutualInfo_refrac_LargeBin_raw}
\end{align}
The first two lines agree with the two-bin case, so that we have
\begin{align}
{\cal I}\left(2\dt\right)-{\cal I}\left(\dt,\dt\right)= & -\dt\left\langle \lambda_{1}+\lambda_{2}\right\rangle _{\boldsymbol{\lambda}}\ln\left(\left\langle \lambda_{1}+\lambda_{2}\right\rangle _{\boldsymbol{\lambda}}\dt\right)\label{eq:Diff_MI_longbin}\\
 & +\dt\left\langle \left(\lambda_{1}+\lambda_{2}\right)\ln\left(\left(\lambda_{1}+\lambda_{2}\right)\dt\right)\right\rangle _{\boldsymbol{\lambda}}\label{eq:Diff_MI_longbin_II}\\
 & -\left[\dt\left(\left\langle \lambda_{1}\right\rangle _{\boldsymbol{\lambda}}\ln\left(\left\langle \lambda_{1}\right\rangle _{\boldsymbol{\lambda}}\dt\right)+\left\langle \lambda_{2}\right\rangle _{\boldsymbol{\lambda}}\ln\left(\left\langle \lambda_{2}\right\rangle _{\boldsymbol{\lambda}}\dt\right)\right)\right.\label{eq:Diff_MI_doublebin}\\
 & \left.+\dt\left(\left\langle \lambda_{1}\ln\left(\lambda_{1}\dt\right)\right\rangle _{\boldsymbol{\lambda}}+\left\langle \lambda_{2}\ln\left(\lambda_{2}\dt\right)\right\rangle _{\boldsymbol{\lambda}}\right)\right].\label{eq:Diff_MI_doublebin_II}
\end{align}
We assume in the following that the firing rates fluctuate weakly
in the sense that 
\[
\lambda_{i}=\lambda_{0}+\delta\lambda_{i},\ \frac{\delta\lambda_{i}}{\lambda_{0}}\ll1,\ i\in\left\{ 1,2\right\} .
\]
We note that this condition is not met for the example we employ in
the main text, for \figref{Toy_model_proper_definition}, for which
the rate switches between two rates, of which one is one order of
magnitude smaller than the other one. If we use the following considerations
to interpret our observations, we therefore have to take them with
a grain of salt, to say the least. We find them instructive nonetheless.

We can expand \eqref{Diff_MI_longbin} and \eqref{Diff_MI_longbin_II}
in small $\delta\lambda$
\begin{align}
 & \dt\left[\left\langle \left(\lambda_{1}+\lambda_{2}\right)\ln\left(\left(\lambda_{1}+\lambda_{2}\right)\dt\right)\right\rangle _{\boldsymbol{\lambda}}-\left\langle \lambda_{1}+\lambda_{2}\right\rangle _{\boldsymbol{\lambda}}\ln\left(\left\langle \lambda_{1}+\lambda_{2}\right\rangle _{\boldsymbol{\lambda}}\dt\right)\right]\nonumber \\
= & \dt\left[\left\langle \left(2\lambda_{0}+\delta\lambda_{1}+\delta\lambda_{2}\right)\ln\left(2\lambda_{0}+\delta\lambda_{1}+\delta\lambda_{2}\right)\right\rangle _{\boldsymbol{\lambda}}-2\lambda_{0}\ln\left(2\lambda_{0}\right)\right]\nonumber \\
= & \dt\left[\left\langle \left(2\lambda_{0}+\delta\lambda_{1}+\delta\lambda_{2}\right)\left[\ln\left(2\lambda_{0}\dt\right)+\ln\left(1+\frac{\delta\lambda_{1}+\delta\lambda_{2}}{2\lambda_{0}}\right)\right]\right\rangle _{\boldsymbol{\lambda}}-2\lambda_{0}\ln\left(2\lambda_{0}\dt\right)\right]\nonumber \\
= & \dt\left\langle \left(\delta\lambda_{1}+\delta\lambda_{2}\right)\left[\left(\frac{\delta\lambda_{1}+\delta\lambda_{2}}{2\lambda_{0}}\right)-2\lambda_{0}\frac{1}{2}\left(\frac{\delta\lambda_{1}+\delta\lambda_{2}}{2\lambda_{0}}\right)^{2}\right]\right\rangle _{\boldsymbol{\lambda}}+\mathcal{O}\left(\left(\frac{\delta\lambda_{i}}{\lambda_{0}}\right)^{3}\right)\nonumber \\
= & \dt\frac{1}{4\lambda_{0}}\left\langle \left(\delta\lambda_{1}+\delta\lambda_{2}\right)^{2}\right\rangle _{\boldsymbol{\lambda}}+\mathcal{O}\left(\left(\frac{\delta\lambda_{i}}{\lambda_{0}}\right)^{3}\right).\label{eq:Rewrite_part_of_MI2dt}
\end{align}
The last two lines of the expression for ${\cal I}\left(2\dt\right)-{\cal I}\left(\dt,\dt\right)$,
\eqref{Diff_MI_doublebin} and \eqref{Diff_MI_doublebin_II}, yield
\begin{align}
 & -2\dt\left(\left\langle \lambda\ln\left(\lambda\,\dt\right)\right\rangle _{\lambda}-\left\langle \lambda\right\rangle _{\lambda}\ln\left(\left\langle \lambda\right\rangle _{\lambda}\dt\right)\right)\nonumber \\
= & -\dt\frac{\left\langle \left(\delta\lambda\right)^{2}\right\rangle _{\lambda}}{\lambda_{0}}+\mathcal{O}\left(\left(\frac{\delta\lambda}{\lambda_{0}}\right)^{3}\right).\label{eq:Rewrite_part_of_dtdt}
\end{align}
We observe that \eqref{Rewrite_part_of_MI2dt} and \eqref{Rewrite_part_of_dtdt}
agree for the case of perfect correlation, otherwise the estimate
from the large time bin is lower than that for two small time bins.
For the difference between the mutual informations in the large time
bin and in the two small time bins, this means
\begin{align*}
 & {\cal I}\left(2\dt\right)-{\cal I}\left(\dt,\dt\right)\\
= & \dt\frac{1}{4\lambda_{0}}\left\langle \left(\delta\lambda_{1}+\delta\lambda_{2}\right)^{2}\right\rangle _{\boldsymbol{\lambda}}-\frac{\dt}{\lambda_{0}}\left(\left\langle \delta\lambda^{2}\right\rangle \right)\\
= & \frac{\dt}{2\lambda_{0}}\left(\left\langle \delta\lambda_{1}\delta\lambda_{2}\right\rangle -\left\langle \delta\lambda^{2}\right\rangle \right)
\end{align*}

In our toy model with the rate of the imhomogeneous Poisson process
switching itself in a Poisson fashion, we can make this more precise.
We determine the correlation of the rate fluctuations by taking into
account that there are only two possibilities: either the rate does
not switch from one time bin to the next - which is true with probability
$\exp\left(-\dt/T_{\mathrm{switch}}\right)$; then the two rate fluctuations
$\delta\lambda_{1}$ and $\delta\lambda_{2}$ are identical, or there
is a switch; then the fluctuations have opposite sign, which happens
with probability $1-\exp\left(-\dt/T_{\mathrm{switch}}\right)$. Therefore,
we have
\begin{align*}
\left\langle \delta\lambda_{1}\delta\lambda_{2}\right\rangle _{\boldsymbol{\lambda}} & =\left\langle \left(\delta\lambda\right)^{2}\right\rangle _{\lambda}\left(2e^{-\frac{\dt}{T_{\mathrm{switch}}}}-1\right)\\
 & =\left\langle \left(\delta\lambda\right)^{2}\right\rangle _{\lambda}\left(1-2\frac{\dt}{T_{\mathrm{switch}}}\right)+\mathcal{O}\left(\dt^{2}\right)
\end{align*}
and thus
\[
{\cal I}\left(2\dt\right)-{\cal I}\left(\dt,\dt\right)=-\frac{\dt^{2}}{\lambda_{0}T_{\mathrm{switch}}}\left\langle \delta\lambda^{2}\right\rangle +\mathcal{O}\left(\dt^{3}\right)
\]
and therefore
\begin{align*}
\MIR\left(2\dt\right)-\MIR\left(\dt,\dt\right) & =\frac{{\cal I}\left(2\dt\right)-{\cal I}\left(\dt,\dt\right)}{2\dt}\\
 & =-\frac{\dt^{2}}{2\lambda_{0}T_{\mathrm{switch}}}\left\langle \delta\lambda^{2}\right\rangle +\mathcal{O}\left(\dt^{2}\right).
\end{align*}
Finally, we put this result together with \eqref{Increase_MIR_double_bins},
using again that $\left\llangle \lambda_{1}\lambda_{2}\right\rrangle =\left\langle \delta\lambda^{2}\right\rangle +\mathcal{O}\left(\dt\right)$
to obtain
\[
\MIR\left(2\dt\right)-\MIR\left(\dt\right)=\frac{1}{2}\dt\left\langle \delta\lambda^{2}\right\rangle \left(1-\frac{1}{\lambda_{0}T_{\mathrm{switch}}}\right)+\mathcal{O}\left(\dt^{2}\right).
\]
So if there is on average at least one spike between two switches
of the firing rate, the estimate for the $\MIR$ will increase as
well by increasing the time-bin size.

\subsection{Adapting the generalized linear model\label{app:Adapting_Mahuas}}

Building on \citet{Mahuas23_024406}, we adapt the parameters of the
generalized linear model (GLM) in order to generate data resembling
the one from experiments. To be precise, we compare the autocorrelations
and the interspike-interval distributions (ISIs) of the artificial
neuron to the corresponding measures from that cell in the the off
population presented in the main text in \figref{Real_data_MainText}c,d,
stimulated by a black-and-white checkerboards. This neuron is quite
typical for the population and seems to be a good example in terms
of data quality, see \figref{Comparison_real_artificial}. We refer
to the supplemental material of \citet{Mahuas23_024406} for a detailed
description of the GLM and paraphrase here only the parts needed to
understand our parameter changes.

The spiking rate in the GLM is here given by
\[
\lambda_{i}\left(t\right)=\frac{1}{1+e^{-h_{i}\left(t\right)}},\ h_{i}\left(t\right)=h_{i}^{\mathrm{bias}}+h_{i}^{\mathrm{stim}}\left(t\right)+h_{i}^{\mathrm{int}}\left(t\right),
\]
where $h_{i}^{\mathrm{stim}}\left(t\right)$ contributes the effect
of the stimulus, $h_{i}^{\mathrm{int}}\left(t\right)$ the self-coupling
of the neuron (mimicking refractoriness) and through $h_{i}^{\mathrm{bias}}$
one controls the basic level activity. $h_{i}^{\mathrm{stim}}\left(t\right)$
is given by the (temporal) convolution of the input with the difference
of two raised cosine functions, as in 
\[
\mathrm{rc}\left(\tau,s,c\right)=\begin{cases}
\cos^{2}\left(\frac{\pi}{2}\left(\ln\left(\tau+s\right)-c\right)\right), & -1\leq\ln\left(\tau+s\right)-c\leq1\\
0, & \mathrm{otherwise}.
\end{cases}
\]
To make the response of the neuron faster, we change the parameters
of \citet{Mahuas23_024406} according to $c_{1}=4.8\rightarrow4.1$,
$c_{2}=5.3\rightarrow4.6$ and $s=50\rightarrow25$, which compresses
the kernel by approximately a factor $2$ (note the exponential relation
between $\tau$ and $c$). Furthermore, we changed the overall prefactor
entering in $h_{i}^{\mathrm{stim}}\left(t\right)$ from $0.5$ to
$2$. Additionally, we change the bias from $-4$ to $-3$, which
increases the mean firing rate. Finally, we replace the absolute refractory
period of $10\,\ms$ by an absolute refractory period of $5\,\ms$,
followed by a relative refractory period with exponential recovery
with a decay time of $10\,\ms$. This yields the statistics shown
in figure 7 (left column).

\begin{figure}
\includegraphics[width=1\textwidth]{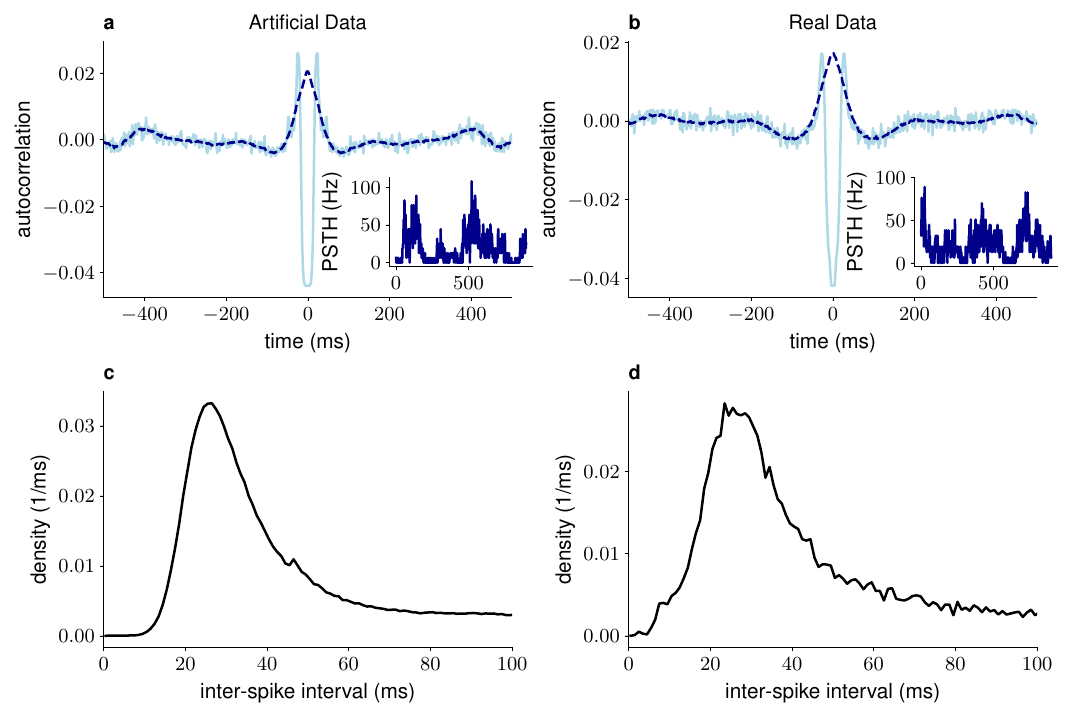}\caption{Comparison of the statistics of artificial and real data. We analyze
the spiking activity of the model described in this to the recordings
presented in \secref{Results_retinal_data}, artificial data in the
left, real data in the right column. The autocorrelation of the PSTH
and the spikes are shown in panels a and b, respective insets show
example section of the PSTH. Inter-spike intervals are shown in panel
c and d.\label{fig:Comparison_real_artificial}}
\end{figure}

\subsection{Computing MIRs for real data in praxis\label{app:RealData_computeMIR}}

\begin{figure}
\includegraphics{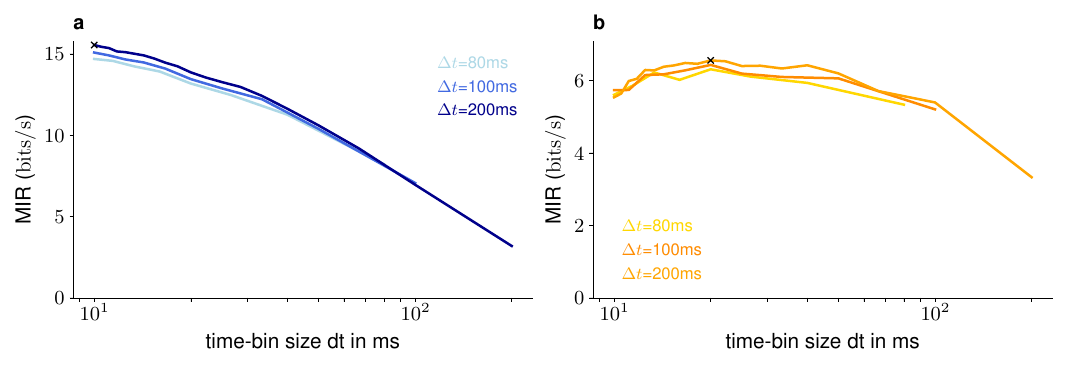}

\caption{\textbf{(a)} The dependence of the $\protect\MIR$ estimate on the
time-bin size for the neuron of \figref{Real_data_MainText}. Final
estimate of the $\protect\MIR$ (with $\protect\dt$ fixed) indicated
by blue cross. \textbf{(d) }Like panel a, but the experiment from
\figref{Comparison_Scatter_StimExpTypes}.\label{fig:MIR_vs_dt_RealData}}
\end{figure}

In this part of the appendix, we explain how to choose appropriate
values for $\dt$ and $\Delta t$ to analyze real data. To achieve
this, we compare the estimate of the $\MIR$ for different values
of $\Delta t$ and varying time-bin size $\dt$ in \figref{MIR_vs_dt_RealData}.
The differently nuanced curves indicate different values of $\Delta t$.
Because in our examples they nearly lie on top of each other, we are
confident that our values $\Delta t$ with which we perform the final
estimate of the $\MIR$ is large enough. 

From the $\dt$-$\MIR$ curve, we can read of what is a good value
for the time bin. We know that it should be decreasing because a lower
time resolution decreases the information. For the neuron whose data
is shown in \figref{MIR_vs_dt_RealData}a, this is the case and we
therefore take the lowest value for $\dt$ for which convergence is
about to be reached. In this experiment, there are 79 repetitions,
quite a lot for a neuroscience experiment, whereas for experiment
whose data is shown \figref{MIR_vs_dt_RealData}b, there are only
54 repetition, which makes it more challenging to analyze. Indeed,
we observe that the corresponding $\dt$-$\MIR$ curve is not monotonous.
This behavior might derive from the bias due to the lower number or
repetitions. Also, it could be that our approximation in the regime
of small time bins is imprecise for this neuron and small time bins.

By using \figref{MIR_vs_dt_RealData}, we can detect this behavior
with our analysis and deal with it. As the best proxy, we choose
the maximum of the curve. By making this choice, we assume that the
true value of $\MIR$ at $\dt\rightarrow0$ does not deviate much
from its equivalent estimate at moderate time-bin size. Note that
this is an assumption that, by its very nature, we cannot check because
it would require to either reduce $\Delta t$, which is prohibitive
because of the correlation time of the stimulus, or to increase the
number of repetitions, which we cannot achieve either, of course.
For some neurons we checked, the failure of full Moba-MIRA is even
clearer than in the example shown in \figref{MIR_vs_dt_RealData},
in one case even leading to negative estimates in the limit of small
$\dt$, shown in figure A3b. Note however, that for all the retinal
recordings we checked the variant of mixed Moba-MIRA yields reasonable
results - without a comparable failure. While it limits the range
of number of time bins (and therefore $\Delta t$) that one can use,
because the histogram method is numerically expensive, this is therefore
still a resort that one can take, given that for our data no huge
$\Delta t$ are necessary. An important use of plots like in \figref{MIR_vs_dt_RealData}
is to check that the approximation leads to reasonable results. 

\subsection{Strategies of removing the bias in estimates\label{sec:Appendix_bias_removal}}

How do we deal with the problem of biases in the estimation of entropies
for high-dimensional data? Indeed, because the entropy is a concave
function of the probability, an imprecise estimate of the latter leads
to underestimating the entropy. Because we have many samples for the
output entropy, the effect of fluctuations is negligible there, whereas
for the input entropy, we only have as many samples as repetitions
of the same stimulus, so at most about $100$, and therefore a much
bigger bias. In the following, we will therefore focus on how to remove
the bias from the input entropy.

Although the bias is reduced by projecting the measured statistics
on a statistical model, it is still large enough to spoil the results
for the typical number of repetitions available in recordings. To
reduce this remaining bias, which is due to suboptimal estimates for
the covariances across time bins, we employ what is known as shuffling
approach \citet{Montemurro07_2913,Mahuas23_024406}: we shuffle the
spike times across repetitions, removing the noise correlations. Conditional
on the stimulus, the activity across time bins should therefore be
independent and the estimate for the entropy equal that one for the
single time bins. However, this will not be the case, precisely because
of the bias explained before. Assuming that this bias is the same
for the shuffled and the unshuffled data, we take the difference of
the resulting entropy from the one of the single-bin estimate to estimate
it. Concretely, we take
\[
S_{\mathrm{across\ bin},\mathrm{de-biased}}^{\mathrm{in}}=S_{\mathrm{across\ bin}}^{\mathrm{in}}-S_{\mathrm{across\ bin,\mathrm{shuffled}}}^{\mathrm{in}}+S_{\mathrm{single}}^{\mathrm{in}}.
\]
In addition, we also reduce the bias in the contribution from the
single bins in a similar way: by shuffling the spike times across
time, we make all time bins statistically equivalent, removing the
variability due to the stimulus. In the shuffled data, each time bin
therefore has the same statistics as the the neural activity sampled
across all time bins. We therefore have
\[
\lim_{N_{\mathrm{rep}}\rightarrow\infty}S_{\mathrm{single},\mathrm{shuffled\ in\ time}}^{\mathrm{in}}\left(N_{\mathrm{rep}}\right)=\lim_{N_{\mathrm{rep}}\rightarrow\infty}S_{\mathrm{single}}^{\mathrm{out}}\left(N_{\mathrm{rep}}\right).
\]
For a finite number of repetitions, however, also $S_{\mathrm{single},\mathrm{shuffled\ in\ time}}^{\mathrm{in}}\left(N_{\mathrm{rep}}\right)$
will be biased. As before, we assume that this bias is the same as
for the unshuffled version and therefore  take
\[
S_{\mathrm{single},\mathrm{de-biased}}^{\mathrm{in}}\left(N_{\mathrm{rep}}\right)=S_{\mathrm{single}}^{\mathrm{in}}\left(N_{\mathrm{rep}}\right)-S_{\mathrm{single},\mathrm{shuffled\ in\ time}}^{\mathrm{in}}\left(N_{\mathrm{rep}}\right)+S_{\mathrm{single}}^{\mathrm{out}}\left(N_{\mathrm{rep}}\right).
\]
This means for the mutual information that we have
\[
{\cal I}_{\mathrm{single,\mathrm{de-biased}}}\left(N_{\mathrm{rep}}\right)=S_{\mathrm{single},\mathrm{shuffled\ in\ time}}^{\mathrm{in}}\left(N_{\mathrm{rep}}\right)-S_{\mathrm{single}}^{\mathrm{in}}\left(N_{\mathrm{rep}}\right).
\]
From figure A3, it is apparent that these additional steps are important
for the good performance of our method, whereas they do not improve,
but rather deteriorate, the estimate according to \citet{Strong98_197},
based on the histogram method. We reckon that this comes about by
the introduction of spurious higher-order correlations by the shuffling,
to which the histogram method is sensitive. They lead to an overestimate
of the bias, rendering the shuffling trick useless for this method.

In addition to the shuffling trick, we also regularize our estimates
of the covariances, by taking
\[
c_{t}^{\mathrm{noise},\,\mathrm{est}}=\left(1-\epsilon\right)c_{t}^{\mathrm{noise}}+\epsilon\bar{c}^{\mathrm{noise}},
\]
where $\bar{c}^{\mathrm{noise}}$ is the time average over all noise
covariances. This is a common procedure to reduce noise in this estimate
and therefore the bias in the entropy \citet{Nejatbakhsh23_Neurips}.
\begin{figure}
\includegraphics[width=1\textwidth]{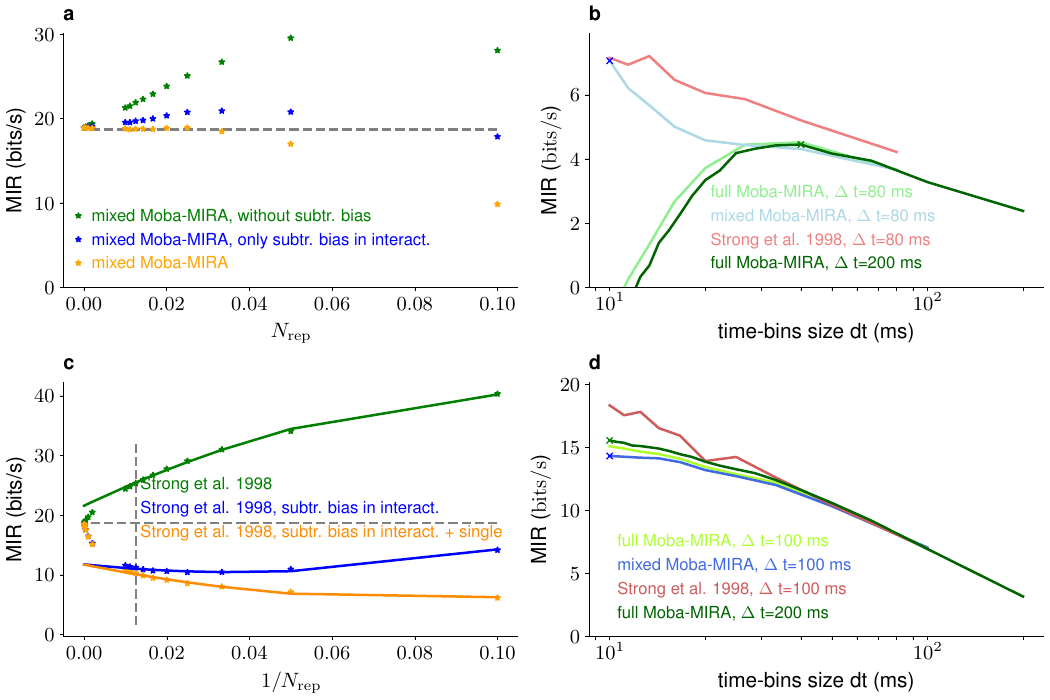} 

\caption{\label{fig:Bias_comparison_methods}\textbf{(a)} Dependence of the
estimate from mixed Moba-MIRA on $N_{\mathrm{rep}}$ for the artificial
data of \figref{Validation_on_ArtData_Overview} and \figref{Validation_on_ArtData_Detail},
$k=10$. \textbf{(b)} Same for the histogram method \citet{Strong98_197},
together with quadratic extrapolation in $\frac{1}{N_{\mathrm{rep}}}$
(fitted for all $N_{\mathrm{rep}}\protect\geq20$). \textbf{(c) }Estimate
of $\protect\MIR$ of an ON cell for different methods from the population
presented in \figref{Comparison_Scatter_StimExpTypes}, panel c. \textbf{(d)
}Same for off cell from the population shown in \figref{Real_data_MainText},
panel c and d.}
\end{figure}

\end{acknowledgments}

\end{document}